\documentclass[pra,twocolumn,superscriptaddress,showpacs,nofootinbibfloatfix,amsmath,amsfonts,amssymb]{revtex4-2}%
\usepackage{amsmath,amsfonts,amssymb,color}
\usepackage{amsthm}
\usepackage{leftidx}
\usepackage{graphicx}
\usepackage{xcolor}
\usepackage{dcolumn}
\usepackage{bm}
\usepackage{epstopdf}
\usepackage{epsfig}
\usepackage{environ}
\usepackage{pdfcomment}

\usepackage{float}
\usepackage[T1]{fontenc}
\usepackage[latin9]{inputenc}
\usepackage{setspace}
\usepackage{esint}

\begin{document}
	
\preprint{APS/123-QED}

\title{Non-Abelian generalization of non-Hermitian quasicrystal: ${\cal PT}$-symmetry
	breaking, localization, entanglement and topological transitions}

\author{Longwen Zhou}
\email{zhoulw13@u.nus.edu}
\affiliation{%
	College of Physics and Optoelectronic Engineering, Ocean University of China, Qingdao, China 266100
}

\date{\today}

\begin{abstract}
Non-Hermitian quasicrystal forms a unique class of matter with 
symmetry-breaking, localization and topological transitions induced by gain
and loss or nonreciprocal effects. In this work, we introduce a non-Abelian
generalization of non-Hermitian quasicrystal, in which the interplay
between non-Hermitian effects and non-Abelian quasiperiodic potentials
create mobility edges and rich transitions among extended,
critical and localized phases. These generic features are demonstrated
by investigating three non-Abelian variants of the non-Hermitian Aubry-Andr\'e-Harper
model. A unified characterization is given to their spectrum, localization,
entanglement and topological properties. Our findings thus add new
members to the family of non-Hermitian quasicrystal and uncover unique
physics that can be triggered by non-Abelian effects in non-Hermitian
systems.
\end{abstract}

\maketitle

\section{Introduction\label{sec:Int}}

Quasicrystals represent a type of aperiodic system with correlated
disorder. They could exhibit metal-insulator transitions at finite
amounts of quasiperiodic modulations even in one spatial dimension
\cite{QC1,QC2,QC3,QC4,QC5}. In recent years, gain and loss or nonreciprocal
effects have been added to quasiperiodic lattices, leading to
the discovery of a wide varieties of non-Hermitian quasicrystals (NHQCs)
\cite{NHQC0,NHQC1,NHQC2,NHQC3,NHQC4,NHQC5,NHQC6,NHQC7,NHQC8,NHQC9,NHQC10,NHQC11,NHQC12,NHQC13,NHQC14,NHQC15,NHQC16,NHQC17,NHQC18,NHQC19,NHQC20,NHQC21,NHQC22,NHQC23,NHQC24,NHQC25,NHQC26,NHQC27,NHQC28,NHQC29,NHQC30,NHQC31,NHQC32,NHQC33,NHQC34,NHQC35,NHQC36,NHQC37,NHQC38,NHQC39}.
In typical situations, an NHQC with ${\cal PT}$-symmetry could possess
a ${\cal PT}$-breaking transition together with a localization
transition with the change of its non-Hermitian control parameters.
This non-Hermitian localization transition might be further accompanied
by the quantized change of a spectral winding number of the system,
endowing it with nontrivial topological features \cite{NHQC3,NHQC4}.
In the presence of long-range hoppings \cite{NHQC6}, sublattice structures
\cite{NHQC21} or time-periodic modulations \cite{NHQC30}, the extended
and localized phases in an NHQC can also be separated by a critical
phase with mobility edges, and reentrant localization transitions
could be generated by non-Hermitian effects in these cases. Despite
great theoretical efforts, phase transitions in NHQCs have also been
observed experimentally in photonic settings \cite{NHQC34,NHQC35},
motivating further studies of their intriguing topological and transport
properties.

In lattice models, the realization and implication of non-Abelian
gauge fields have attracted continued interest in the quantum engineering
of materials and artificial systems (see Refs.~\cite{NAGPRev1,NAGPRev2,NAGPRev3,NAGPRev4,NAGPRev5,NAGPRev6,NAGPRev7,NAGPRev8}
for reviews). Recently, genuine non-Abelian conditions were considered
in different types of Hofstadter-like models \cite{Hofstadter}, where
intriguing forms of butterfly fractal spectra, localization transitions
and gapped or gapless topological phases were identified \cite{NAHH1,NAHH2,NAHH3,NAHH4,NAHH5,NAHH6,NAHH7,NAHH8,NAHH9,NAHH10,NAHH11,NAHH12,NAHH13,NAHH14}.
Among them, the interplay between non-Hermiticity and non-Abelian
gauge potential has been found to be able to generate trivial, quantum
spin Hall and metallic phases in superconducting circuits \cite{NAHH9}.
Metal-insulator transitions in quasiperiodic lattices with non-Abelian
U(2) and SU(2) gauge fields were also explored in several studies
\cite{NAHH2,NAHH11}. However, much less is known about what happens
when non-Hermitian and non-Abelian effects coexist in a quasicrystal.
Specially, it is interesting to know whether the cooperation between
non-Hermiticity and non-Abelian quasiperiodic potentials could generate
unique phases and transitions that are absent in the Hermitian or
Abelian counterparts of the underlying systems. The resolution of
this issue is also of great experimental relevance due to the recent
realizations of NHQCs by photonic quantum walks, where synthetic spin-$1/2$
degrees of freedom and SU(2) gauge structures could appear
in the dynamics \cite{NHQC34,NHQC35}. These settings thus provide
natural platforms to detect the potential new features of non-Abelian NHQCs.

In this work, we explore the combined effects of non-Hermiticity and
non-Abelian potentials in one-dimensional (1D) quasicrystals. Focusing
on prototypical settings of non-Hermitian Aubry-André-Harper (NHAAH)
models, we reveal that the presence of non-Abelian quasiperiodic superlattices
could extend a critical point of non-Hermitian localization into a
critical region, in which extended and localized eigenstates coexist
and are separated by mobility edges. With the change of gain and loss
or nonreciprocal hopping strengths, the system could further enter
and leave this non-Abelian-effect-induced critical phase through two
different localization transitions, which each of them being characterized
by the quantized jump of a spectral topological winding number.
These properties are expected to be generic for any 1D non-Abelian NHQCs.
In Secs.~\ref{sec:MM} and \ref{sec:MS}, we introduce our non-Abelian NHAAH models, discuss
their non-Abelian conditions, analyze their ${\cal PT}$-symmetries
and present our methods of investigating their eigenspectrum, inverse
participation ratios (IPRs), entanglement spectrum (ES), entanglement
entropy (EE) and topological invariants. In Sec.~\ref{sec:Res}, we
uncover the phases and transitions in our three representative non-Abelian
NHQC models with different types of non-Hermitian effects. The physical properties
of all these models are found to be much richer than their Abelian or Hermitian
counterparts. 
They also form prototypical setups for the theoretical and
experimental study of other non-Abelian potentials in NHQCs.
We summarize our results and discuss potential future
studies in Sec.~\ref{sec:Sum}.

\section{Model}\label{sec:MM}

In this work, we focus on non-Abelian generalizations of the NHAAH
model. The latter represents a typical system in the study of NHQC.
The Hamiltonian of an Abelian NHAAH model can usually be defined as
\begin{alignat}{1}
	H_{0} = \sum_{n}&[J_{{\rm L}}c_{n}^{\dagger}c_{n+1}+J_{{\rm R}}c_{n+1}^{\dagger}c_{n}\nonumber\\
	+ & V\cos(2\pi\alpha n+i\gamma)c_{n}^{\dagger}c_{n}].\label{eq:H0}
\end{alignat}
Here $c_{n}^{\dagger}$ ($c_{n}$) is the creation (annihilation)
operator of a particle on the lattice site $n$. $J_{{\rm L}}$ ($J_{{\rm R}}$)
represents the nearest-neighbor hopping amplitude of a particle from
right to left (left to right) in the lattice. $V$ is the amplitude
of the superlattice onsite potential. $i\gamma$ represents an imaginary
phase factor for $\gamma\in\mathbb{R}\backslash\{0\}$, which could
introduce onsite gain and loss into the lattice. The system described by
$H_{0}$ becomes a 1D quasicrystal once $\alpha$
is chosen to be an irrational number. When $J_{{\rm L}}=J_{{\rm R}}=J$,
it was shown that the system could undergo a ${\cal PT}$ transition,
a localization transition and a topological transition all at once
under the periodic boundary condition (PBC) if $e^{|\gamma_{c}|}=|2J/V|$
\cite{NHQC4}. When $|\gamma|<|\gamma_{c}|$, all the eigenstates
of $H_{0}$ are spatially extended with real eigenvalues, and
the spectral topological winding number $w=0$. When $|\gamma|>|\gamma_{c}|$,
all the eigenstates of $H_{0}$ are spatially localized with complex
eigenvalues, and the spectral topological winding number $w=1$. 
When $J_{{\rm L}}\neq J_{{\rm R}}^{*}$
and $\gamma=0$, $H_{0}$ describes an NHQC with nonreciprocal hoppings.
In this case, a ${\cal PT}$, localization and topological triple
phase transition was also found in the system under the PBC \cite{NHQC3}.
Assuming $J_{{\rm L}}=Je^{-\beta}$ and $J_{{\rm R}}=Je^{\beta}$
($\beta\in\mathbb{R}$), the transition point is determined by $e^{|\beta_{c}|}=|V/(2J)|$.
When $|\beta|<|\beta_{c}|$, all the eigenstates are localized with
real eigenvalues, and the spectral topological winding number $w=0$.
When $|\beta|>|\beta_{c}|$, all the eigenstates are extended with
complex eigenvalues, and the spectral topological winding number $w=1$.
Such a triple phase transition was recently observed in nonunitary
quantum walks \cite{NHQC34}.
It is not hard to see that these two Abelian NHAAH models are dual
with each other under the PBC. Up to the rescaling of some system
parameters, one of them can be mapped to the other under discrete
Fourier transformations \cite{NHQC3,NHQC4}. Meanwhile, no critical
phases with mobility edges were found in these Abelian NHQCs.

We now introduce an SU(2) extension of $H_{0}$, which serves our
purpose of studying non-Abelian NHQCs. Based on the necessary and
sufficient condition of non-Abelian potential for Aubry-Andr\'e-Harper (AAH) models \cite{NAHH12},
we define the Hamiltonian of our system as
\begin{alignat}{1}
	H=\sum_{n}&[J_{{\rm L}}{\bf c}_{n}^{\dagger}\sigma_{0}{\bf c}_{n+1}+J_{{\rm R}}{\bf c}_{n+1}^{\dagger}\sigma_{0}{\bf c}_{n}\nonumber\\
	+ & V{\bf c}_{n}^{\dagger}(e^{-i\phi}\Theta_{n}+e^{i\phi}\Theta_{n}^{-1}){\bf c}_{n}].\label{eq:H}
\end{alignat}
Here, different from $H_{0}$, ${\bf c}_{n}^{\dagger}=(c_{n,\uparrow}^{\dagger},c_{n,\downarrow}^{\dagger})$
denotes a two-component creation operator. $c_{n,\sigma}^{\dagger}$
creates a particle with spin $\sigma$ ($=\uparrow,\downarrow$) on
the lattice site $n$. $\sigma_{0}$ is the two by two identity matrix
acting on the two spin degrees of freedom of the particle. $\phi\in[-\pi,\pi)$
represents an Abelian phase factor. $V$ represents the amplitude
of onsite potential. In this work, we choose 
\begin{equation}
	\Theta_{n}=e^{i\theta_{n}\sigma_{y}}e^{i\theta_{n}\sigma_{z}},\label{eq:Theta}
\end{equation}
where $\sigma_{y}$ and $\sigma_{z}$ are Pauli matrices. This choice
could realize the effect of a genuine non-Abelian gauge potential
if $[e^{i\theta_{n}\sigma_{y}}e^{i\theta_{n}\sigma_{z}},e^{i\theta_{n}\sigma_{z}}e^{i\theta_{n}\sigma_{y}}]\neq0$,
which can be satisfied for $\theta_{n}\neq\{0,\pi/2,\pi,3\pi/2\}$
\cite{NAHH12}. For the models considered in this work, the choices
of phase modulation $\theta_{n}$ are summarized in Table \ref{tab:Mod}.
The non-Abelian condition is clearly met when the $\alpha$ in Table \ref{tab:Mod} takes irrational
values, which is also required for us
to realize a quasiperiodic potential. Note that the form
of non-Abelian potential $\Theta_{n}$ is not unique, and other types
of non-Abelian Hermitian AAH models exist in the literature \cite{NAHH1,NAHH2,NAHH11}.
Nevertheless, the models considered in this study should be sufficient
to capture the general properties originated from the interplay between
non-Hermitian and non-Abelian effects in 1D NHQCs.

\begin{table}
	\begin{centering}
		\begin{tabular}{|c|c|c|}
			\hline 
			Model & Hopping amplitude & Phase modulation\tabularnewline
			\hline 
			\hline 
			1 & $J_{{\rm L}}=J,\quad J_{{\rm R}}=0$ & $\theta_{n}=2\pi\alpha n$\tabularnewline
			\hline 
			2 & $J_{{\rm L}}=Je^{-\beta},\quad J_{{\rm R}}=Je^{\beta}$ & $\theta_{n}=2\pi\alpha n$\tabularnewline
			\hline 
			3 & $J_{{\rm L}}=J_{{\rm R}}=J$ & $\theta_{n}=2\pi\alpha n+i\gamma$\tabularnewline
			\hline 
		\end{tabular}
		\par\end{centering}
	\caption{Definitions of model parameters in the Eqs.~(\ref{eq:H}) and (\ref{eq:Theta}).
		$J\in\mathbb{R}$ is the uniform part of hopping amplitude. $\beta\in\mathbb{R}$
		controls the asymmetry of hopping amplitude. $\gamma\in\mathbb{R}$
		controls the non-Hermiticity of onsite potential. $n\in\mathbb{Z}$
		represents the unit cell index. $\alpha$ belongs to $\mathbb{R}\backslash\mathbb{Q}$
		for a quasicrystal. We set $\alpha=(\sqrt{5}-1)/2$
		in this work. Other irrational values of $\alpha$ will generate similar
		results. \label{tab:Mod}}
\end{table}

To investigate the spectral and localization properties of non-Abelian
NHQCs, we need to solve the eigenvalue equation $H|\psi\rangle=E|\psi\rangle$ for our models.
We can first expand the state as $|\psi\rangle=\sum_{n}{\bf c}_{n}^{\dagger}\boldsymbol{\psi}_{n}|0\rangle$,
with $\boldsymbol{\psi}_{n}=(\psi_{n,\uparrow},\psi_{n,\downarrow})^{{\rm T}}$
being a column vector for the two spin components of $|\psi\rangle$
on the site $n$. Inserting the expanded state into the eigenvalue
equation, we arrive at
\begin{equation}
	J_{{\rm L}}\boldsymbol{\psi}_{n+1}+J_{{\rm R}}\boldsymbol{\psi}_{n-1}+V(e^{-i\phi}\Theta_{n}+e^{i\phi}\Theta_{n}^{-1})\boldsymbol{\psi}_{n}=E\boldsymbol{\psi}_{n}.\label{eq:Seq}
\end{equation}
For a lattice with $L$ sites, we denote the eigenenergies of $H$ by $\{E_{j}|j=1,2,...,2L-1,2L\}$.
The right eigenvector of $H$ with energy $E_{j}$ is denoted by $|\psi_{j}\rangle=\sum_{n=1}^{L}\sum_{\sigma=\uparrow,\downarrow}\psi_{n,\sigma}^{j}|n,\sigma\rangle$,
where $|n,\sigma\rangle=c_{n,\sigma}^{\dagger}|\emptyset\rangle$ and $|\emptyset\rangle$ represents the vacuum state. We assume
that each right eigenvector $|\psi_{j}\rangle$ has been properly
normalized, such that $\sum_{n,\sigma}|\psi_{n,\sigma}^{j}|^{2}=1$.
For $J_{{\rm L}}\neq J_{{\rm R}}^{*}$ or $\Theta_{n}^{-1}\neq\Theta_{n}^{\dagger}$,
$H$ is non-Hermitian and its eigenvalues can be complex.

The Models 1--3 defined through the Eqs.~(\ref{eq:H}), (\ref{eq:Theta})
and Table \ref{tab:Mod} could all possess real spectra due to their
${\cal PT}$ symmetries. This can be proved as follows. We first work
out the term $e^{-i\phi}\Theta_{n}+e^{i\phi}\Theta_{n}^{-1}$ in 
Eq.~(\ref{eq:Seq}) by inserting Eq.~(\ref{eq:Theta}) into it as
\begin{alignat}{1}
	& e^{-i\phi}e^{i\theta_{n}\sigma_{y}}e^{i\theta_{n}\sigma_{z}}+e^{i\phi}e^{-i\theta_{n}\sigma_{z}}e^{-i\theta_{n}\sigma_{y}}\nonumber \\
	= & d_{n}^{0}\sigma_{0}+d_{n}^{x}\sigma_{x}+d_{n}^{y}\sigma_{y}+d_{n}^{z}\sigma_{z},\label{eq:PT0}
\end{alignat}
where
\begin{alignat}{1}
	d_{n}^{0} &= \cos\phi\left(\cos2\theta_{n}+1\right),\nonumber \\
	d_{n}^{x} &= \sin\phi\left(\cos2\theta_{n}-1\right),\nonumber \\
	d_{n}^{y} &= d_{n}^{z}=\sin\phi\sin2\theta_{n}.\label{eq:d}
\end{alignat}
The ${\cal P}$ operator leads to the spatial inversion $n\rightarrow-n$
in one dimension and the time-reversal ${\cal T}\equiv{\cal U}{\cal K}$,
where ${\cal U}$ is a unitary operator and ${\cal K}$ takes the
complex conjugation. For $\theta_{n}=2\pi\alpha n+i\gamma$, the combined
action of ${\cal P}$ and ${\cal K}$ leads to $(d_{-n}^{0}\sigma_{0})^{*}=d_{n}^{0}\sigma_{0}$,
$(d_{-n}^{x}\sigma_{x})^{*}=d_{n}^{x}\sigma_{x}$, $(d_{-n}^{y}\sigma_{y})^{*}=d_{n}^{y}\sigma_{y}$
and $(d_{-n}^{z}\sigma_{z})^{*}=-d_{n}^{z}\sigma_{z}$. Since $d_{n}^{y}=d_{n}^{z}$
in Eq.~(\ref{eq:d}), the term $e^{-i\phi}\Theta_{n}+e^{i\phi}\Theta_{n}^{-1}$
is invariant under the combined ${\cal PT}$ operation if ${\cal U}$
could execute a rotation following which $\sigma_{y}\rightarrow\sigma_{z}$
and $\sigma_{z}\rightarrow-\sigma_{y}$ in Eq.~(\ref{eq:PT0}). This
can be achieved by setting ${\cal U}=e^{-i\frac{\pi}{4}\sigma_{x}}$,
so that ${\cal U}\sigma_{y}{\cal U}^{\dagger}=\sigma_{z}$ and ${\cal U}\sigma_{z}{\cal U}^{\dagger}=-\sigma_{y}$.
Therefore, the Hamiltonian of Model 3 is symmetric under the combined
operation of parity ${\cal P}:n\rightarrow-n$ and time-reversal ${\cal T}=e^{-i\frac{\pi}{4}\sigma_{x}}{\cal K}$,
implying that it could have a real spectrum in its ${\cal PT}$-invariant
phase. To see that the Models 1 and 2 also share the same ${\cal PT}$-symmetry
with the Model 3, we take the PBC in Eq.~(\ref{eq:Seq}) and perform
the Fourier transformation for $\boldsymbol{\psi}_{n}$ from position
to momentum spaces as
\begin{equation}
	\boldsymbol{\psi}_{n}=\frac{1}{\sqrt{L}}\sum_{\ell=1}^{L}\boldsymbol{\varphi}_{\ell}e^{-i2\pi\alpha\ell n}.\label{eq:FT}
\end{equation}
The eigenvalue equation for the first two models are then transformed
to
\begin{equation}
	\Lambda_{\ell}\boldsymbol{\varphi}_{\ell}+\Xi\boldsymbol{\varphi}_{\ell+2}+\Xi^{\dagger}\boldsymbol{\varphi}_{\ell-2}=E\boldsymbol{\varphi}_{\ell},\label{eq:Seq2}
\end{equation}
where
\begin{alignat}{1}
	\Lambda_{\ell}= & \left(J_{L}e^{-i2\pi\alpha\ell}+J_{R}e^{i2\pi\alpha\ell}+V\cos\phi\right)\sigma_{0}-V\sin\phi\sigma_{x},\nonumber \\
	\Xi= & \frac{V}{2}\cos\phi\sigma_{0}+\frac{V}{2}\sin\phi\sigma_{x}+\frac{V}{2i}\sin\phi\sigma_{y}+\frac{V}{2i}\sin\phi\sigma_{z}.\label{eq:Seq3}
\end{alignat}
Under the combined actions of ${\cal P}:\ell\rightarrow-\ell$
and ${\cal T}=e^{-i\frac{\pi}{4}\sigma_{x}}{\cal K}$ in momentum
space, we can directly see that ${\cal PT}\Lambda_{\ell}=\Lambda_{\ell}{\cal PT}$
and ${\cal PT}\Xi^{(\dagger)}=\Xi^{(\dagger)}{\cal PT}$. The Hamiltonians
of Models 1 and 2 in momentum representations are thus ${\cal PT}$-symmetric.
Therefore, our Models 1-3 all have the $\cal{PT}$-symmetry. Their spectra
are allowed to be real in the ${\cal PT}$-invariant regions. They may also undergo
$\cal{PT}$-breaking transitions with the change of their non-Hermitian parameters.
The models we introduced thus provide typical settings to explore $\cal{PT}$
transitions in non-Abelian NHQCs.

\section{Method\label{sec:MS}}
In this section, we introduce relevant tools to characterize the spectrum,
localization, entanglement and topological transitions in non-Abelian NHQCs.
To capture the spectral transition of $H$ from real to
complex (or vice versa), we introduce the following quantities
\begin{equation}
	E_{{\rm I}}^{\max}=\max_{j\in\{1,...,2L\}}(|{\rm Im}E_{j}|),\label{eq:EImax}
\end{equation}
\begin{equation}
	E_{{\rm I}}^{\min}=\min_{j\in\{1,...,2L\}}(|{\rm Im}E_{j}|),\label{eq:EImin}
\end{equation}
\begin{equation}
	\rho=\frac{1}{2L}\sum_{j=1}^{2L}\theta(|{\rm Im}E_{j}|),\label{eq:DOS}
\end{equation}
where the step function $\theta(x)=1$ if $x>0$ and $\theta(x)=0$
if $x\leq0$. For a given set of system parameters, we have $E_{{\rm I}}^{\max}=\rho=0$
if all the eigenvalues of $H$ are real. Conversely, if $E_{{\rm I}}^{\min}>0$
and $\rho=1$, all the eigenvalues of $H$ are complex. When $0<\rho<1$,
real and complex eigenvalues of $H$ coexist in the spectrum. Therefore,
we can use the functions in Eqs.~(\ref{eq:EImax})--(\ref{eq:DOS})
to characterize the global spectral properties of $H$ and distinguish
its different spectral regions, i.e., the entirely real, entirely complex,
or a mixture of real and complex eigenvalues.

For the eigenstate $|\psi_{j}\rangle$ of $H$, we can define its
IPRs and normalized participation ratios
(NPRs) as ${\rm IPR}_{j}=\sum_{n,\sigma}|\psi_{n,\sigma}^{j}|^{4}$
and ${\rm NPR}_{j}=(2L\cdot{\rm IPR}_{j})^{-1}$, where the factor $2$
comes from the two spin degrees of freedom. At a given set of system
parameters, the averages of IPRs and NPRs
over all eigenstates are given by $\langle{\rm IPR}\rangle=\frac{1}{2L}\sum_{j=1}^{2L}{\rm IPR}_{j}$
and $\langle{\rm NPR}\rangle=\frac{1}{2L}\sum_{j=1}^{2L}{\rm NPR}_{j}$.
Using these definitions, we can introduce the following quantities
to characterize the localization nature of states in the system
\begin{equation}
	{\rm IPR}^{\max}=\max_{j\in\{1,...,2L\}}({\rm IPR}_{j}),\label{eq:IPRmax}
\end{equation}
\begin{equation}
	{\rm IPR}^{\min}=\min_{j\in\{1,...,2L\}}({\rm IPR}_{j}),\label{eq:IPRmin}
\end{equation}
\begin{equation}
	\eta=\log_{10}(\langle{\rm IPR}\rangle\langle{\rm NPR}\rangle).\label{eq:ETA}
\end{equation}
For an extended (a localized) state, we have ${\rm IPR}_{j}\sim L^{-1}$
(${\rm IPR}_{j}\sim\lambda_{j}$) and ${\rm NPR}_{j}\sim1$ (${\rm NPR}_{j}\sim L^{-1}$)
in the limit $L\rightarrow\infty$, where $\lambda_{j}$ is the $L$-independent
localization length of state $|\psi_{j}\rangle$ \cite{MCMP}. Therefore,
for a given set of system parameters, we have ${\rm IPR}^{\max}\rightarrow0$
if all the eigenstates of $H$ are extended (a metallic phase) and
${\rm IPR}^{\min}>0$ if all the eigenstates of $H$ are localized
(an insulating phase) in the thermodynamic limit. The quantity $\eta$
was first introduced to characterize critical phases, in which extended
and localized eigenstates coexist and are separated in their energies
by mobility edges \cite{Relocal}. Its applicability to NHQCs
was further demonstrated in Ref.~\cite{NHQC29}. Due to the scaling properties
of IPRs and NPRs with the system size, we would have $\eta\sim-\log_{10}(2L)$
for our non-Abelian NHQCs in the extended and localized phases, with
$\eta\rightarrow-\infty$ in both cases in the limit $L\rightarrow\infty$.
In the critical phase, $\langle{\rm IPR}\rangle\langle{\rm NPR}\rangle$
takes a finite value in the thermodynamic limit. $\eta$ can
thus be used to distinguish critical from metallic and insulating
phases. We will use the Eqs.~(\ref{eq:IPRmax})--(\ref{eq:ETA})
to characterize phases with different localization properties and
explore localization-delocalization transitions in our non-Abelian
NHQC models.

In Abelian NHQCs, it has been found that ${\cal PT}$-breaking and
localization transitions are usually accompanied by topological
transitions, which are depicted by quantized jumps of spectral
topological winding numbers \cite{NHQC3,NHQC4}. To check whether
similar results hold in non-Abelian NHQCs, we introduce the definition
of spectral winding numbers under the PBC as
\begin{equation}
	w_{\ell}=\int_{0}^{2\pi}\frac{d\vartheta}{2\pi i}\frac{\partial}{\partial\vartheta}\ln\{\det[H(\vartheta)-{\cal E}_{\ell}]\},\quad\ell=1,2.\label{eq:w12}
\end{equation}
Depending on the explicit form of the model, $\vartheta$ may be interpreted
as a synthetic flux going through the 1D ring formed by the PBC lattice,
or a phase shift in the quasiperiodic superlattice potential. For
the Models 1 and 2 in Table \ref{tab:Mod}, the $\vartheta$-dependence
of $H$ can be introduced by setting $J_{{\rm L}}\rightarrow J_{{\rm L}}e^{-i\vartheta/L}$
and $J_{{\rm R}}\rightarrow J_{{\rm R}}e^{i\vartheta/L}$. For the
Model 3, we can set $2\pi\alpha n+i\gamma\rightarrow2\pi\alpha n+i\gamma+\vartheta/L$
in $\theta_{n}$. The $w_{\ell}$ then counts the number of
times the spectrum of $H(\vartheta)$ winds around a chosen base energy
${\cal E}_{\ell}$ on the complex plane when $\vartheta$ changes over a period from zero
to $2\pi$ \cite{NHQC29}. The base energy ${\cal E}_{\ell}$ is in
general model-dependent. In our numerical calculations, we choose
${\cal E}_{1}$ (${\cal E}_{2}$) to be the real part of energy of
the first (last) eigenstate of $H$ whose IPR deflects from zero with
the change of system parameters. For Abelian NHAAH models with only
extended and localized phases, we expect ${\cal E}_{1}={\cal E}_{2}$
and there is only a single winding number $w=w_{1}=w_{2}$, which
is the case in Refs.~\cite{NHQC3,NHQC4}. For a system that could
also possess critical mobility edge phases, we expect a quantized
jump in $w_{1}$ ($w_{2}$) when the system goes through a transition
point between the extended and critical (critical and localized) phases.
If the system only holds critical and extended (localized) phases,
$w_{2}$ ($w_{1}$) will be ill-defined and we only expect a quantized
change in $w_{1}$ ($w_{2}$) when the system passes through the boundary
between the two phases. Therefore, the winding numbers $(w_{1},w_{2})$
could provide us with sufficient information to describe the topological
nature of localization transitions in non-Abelian NHQCs. Note in passing that
in the Hermitian limit of $H$, $H(\vartheta)$ is also Hermitian
with a real spectrum. $(w_{1},w_{2})$ will then become identically
zero, implying that the definition of winding number $w_{\ell}$ in
Eq.~(\ref{eq:w12}) is unique to non-Hermitian systems. In the meantime,
it is better to interpret the quantized jumps of $(w_{1},w_{2})$
as invariants accompanying topological localization transitions in
NHQCs, instead of treating them as topological numbers characterizing
a whole extended, localized or critical mobility edge phase.

The ES and EE could provide us with important information about topological
and quantum phase transitions from an information-theoretical perspective.
In the context of Abelian NHQCs, localization transitions and mobility
edges have been identified from the ES and EE \cite{NHQC39}. In the
absence of mobility edges, the EE was found to change discontinuously
when the system goes through a localization-delocalization transition.
In the critical phase, the EE of extended and localized
states were found to show clear distinctions around the mobility edge
\cite{NHQC39}. For our non-Abelian NHQCs, we introduce the single-particle
biorthogonal eigenvectors $\{|\psi_{j}^{R}\rangle\}$ and $|\psi_{j}^{L}\rangle$
of $H$ as
\begin{equation}
	H|\psi_{j}^{R}\rangle=E_{j}|\psi_{j}^{R}\rangle,\qquad H|\psi_{j}^{L}\rangle=E_{j}^{*}|\psi_{j}^{L}\rangle,\label{eq:LRvec}
\end{equation}
where $j=1,...,2L$. The bi-normalization condition $\langle\psi_{j}^{L}|\psi_{j'}^{R}\rangle=\delta_{jj'}$
is satisfied and the completeness relation is expressed as $\sum_{j}|\psi_{j}^{R}\rangle\langle\psi_{j}^{L}|=1$.
Assuming that a collection of energy levels $\{E_{m}\}\subseteq\{E_{j}|j=1,...,2L\}$
is populated each by a fermion, the many-particle wavefunction of
the system can be expressed as $|\Psi^{R}\rangle=\prod_{m}\psi_{Rm}^{\dagger}|\emptyset\rangle$
and $|\Psi^{L}\rangle=\prod_{m}\psi_{Lm}^{\dagger}|\emptyset\rangle$,
where $\psi_{Rm}^{\dagger}$ ($\psi_{Lm}^{\dagger}$) creates a single-particle
eigenstate $|\psi_{m}^{R}\rangle$ ($|\psi_{m}^{L}\rangle$) whose energy belongs
to the set $\{E_{m}\}$. The
many-particle density matrix of the system then takes the form $\rho=|\Psi^{R}\rangle\langle\Psi^{L}|$.
To investigate the bipartite entanglement, we consider a partition
of the system into two equal parts A and B in real space \cite{FloqESEE}.
Tracing out the degrees of freedom belonging to the subsystem B, we
obtain the reduced density matrix of subsystem A as $\rho_{{\rm A}}={\rm Tr}_{{\rm B}}\rho$,
and the EE is given by $S=-{\rm Tr}(\rho_{{\rm A}}\ln\rho_{{\rm A}})$
\cite{Peschel2003}. For noninteracting fermions, $\rho_{{\rm A}}$
represents a Gaussian state and we can express it as $\rho_{{\rm A}}=\frac{1}{Z}e^{-H_{{\rm A}}}$,
where $Z$ is a normalization factor. The eigenspectrum $\{\xi_{j}|j=1,...,L\}$
of the entanglement Hamiltonian $H_{{\rm A}}$ forms the ES, which
can be related to the eigenvalues $\{\zeta_{j}|j=1,...,L\}$ of the
single-particle correlation matrix $C$ as $\xi_{j}=\ln(\zeta_{j}^{-1}-1)$
\cite{NHESEE1,NHESEE2,NHESEE3,NHESEE4,NHESEE5,NHESEE6,NHESEE7,NHESEE8}.
Here the matrix elements of $C$ in real-space take the form of
\begin{equation}
	C_{nn'}=\langle\Psi^{L}|c_{n}^{\dagger}c_{n'}|\Psi^{R}\rangle=\langle n'|P|n\rangle,\label{eq:C}
\end{equation}
where $n,n'\in{\rm A}$. The projector $P=\sum_{m}|\psi_{m}^{R}\rangle\langle\psi_{m}^{L}|$,
and the sum is taken over all occupied states. Due to the one-to-one
correspondence between the spectra of $C$ and $H_{{\rm A}}$, we
will also refer to the spectrum $\{\zeta_{j}|j=1,...,L\}$ of the correlation
matrix $C$ as the ES. Meanwhile, the EE can be expressed in
terms of the correlation matrix spectrum as
\begin{equation}
	S=-\sum_{j=1}^{L}[\zeta_{j}\ln\zeta_{j}+(1-\zeta_{j})\ln(1-\zeta_{j})].\label{eq:EE}
\end{equation}
Therefore, both the ES and EE can be obtained from the spectrum of correlation matrix
$C$ restricted to the subsystem A. The tools introduced in this section allow
us to investigate the physics of non-Abelian NHQCs from different and complementary
perspectives. We expect these tools to be applicable to non-Abelian NHQC models beyond those
considered in this work.

\section{Results\label{sec:Res}}

In this section, we reveal the spectrum, localization, entanglement
and topological properties of non-Abelian NHQCs with the methods presented
in the last section. We start with a relatively simple case, in which
the non-Hermiticity is introduced by allowing particles to hop along
only one direction of the lattice (the Model 1 in Table \ref{tab:Mod}).
We then consider two more general situations, in which the non-Hermitian
effects are originated from asymmetric hoppings (the Model 2 in Table
\ref{tab:Mod}) and complex non-Abelian onsite potentials (the Model 3 in Table
\ref{tab:Mod}). Rich connections between different physical properties
of non-Abelian NHQCs will be established for all the cases.

\subsection{Unidirectional hopping\label{subsec:M1}}

We start with a non-Hermitian generalization of non-Abelian AAH model,
in which the hopping terms only allow particles to jump from right
to left between neighboring sites. According to the Eqs.~(\ref{eq:Theta}),
(\ref{eq:Seq}) and Table \ref{tab:Mod}, the eigenvalue equation
of this Model 1 takes the form
\begin{equation}
	J\boldsymbol{\psi}_{n+1}+V(e^{-i\phi}\Theta_{n}+e^{i\phi}\Theta_{n}^{-1})\boldsymbol{\psi}_{n}=E\boldsymbol{\psi}_{n},\label{eq:SeqM1}
\end{equation}
where $\Theta_{n}=e^{i2\pi\alpha n\sigma_{y}}e^{i2\pi\alpha n\sigma_{z}}$.
Referring to the last section, the system described by Eq.~(\ref{eq:SeqM1})
possesses the ${\cal PT}$-symmetry, which implies that it could have
real spectra under the PBC. This should happen when the unidirectional
hopping amplitude $J$, which induces the non-Hermitian effect, is
small compared with the other energy scales of the system. With the
increase of $J$, we expect a transformation of the spectrum from
real to complex when $J$ goes beyond a critical value, which characterizes
the ${\cal PT}$-breaking transition of the system. In the meantime,
when $J\rightarrow0$, the state profiles in the system are controlled
by the non-Abelian onsite potential, and we expect all the eigenstates
to be localized for an irrational $\alpha$. In the opposite limit
$J\rightarrow\infty$, the structures of states are dominated by the
unidirectional hopping term. We expect all the eigenstates to be extended
in this limit for a finite $V$. In between these two limits, there
must be a delocalization transition for the eigenstates with the increase
of $J$. In the Abelian case, such a transition happens at $J_{c}=V/2$
\cite{NHQC12}. All the eigenstates change from localized to extended
when $J$ goes from below to above $J_{c}$, with their eigenvalues
changing from real to complex. The localization and ${\cal PT}$-breaking
transitions in the corresponding Abelian model thus go hand in hand
with each other \cite{NHQC12}. In Eq.~(\ref{eq:SeqM1}), the non-Abelian
potential introduces internal structures to each lattice site, which
may transform the critical point $J_{c}$ into a critical region $J\in(J_{c1},J_{c2})$
with coexisting extended and localized eigenstates that are separated
by mobility edges. The presence of such a critical phase in our system,
which is due to the interplay between non-Hermitian and non-Abelian
effects, will be revealed in the following calculations. Besides,
the ${\cal PT}$ transition point shall be modified and only parts
of eigenvalues in the spectrum may become complex after the transition
first happens, as will be discussed below.

\begin{figure}
	\begin{centering}
		\includegraphics[scale=0.5]{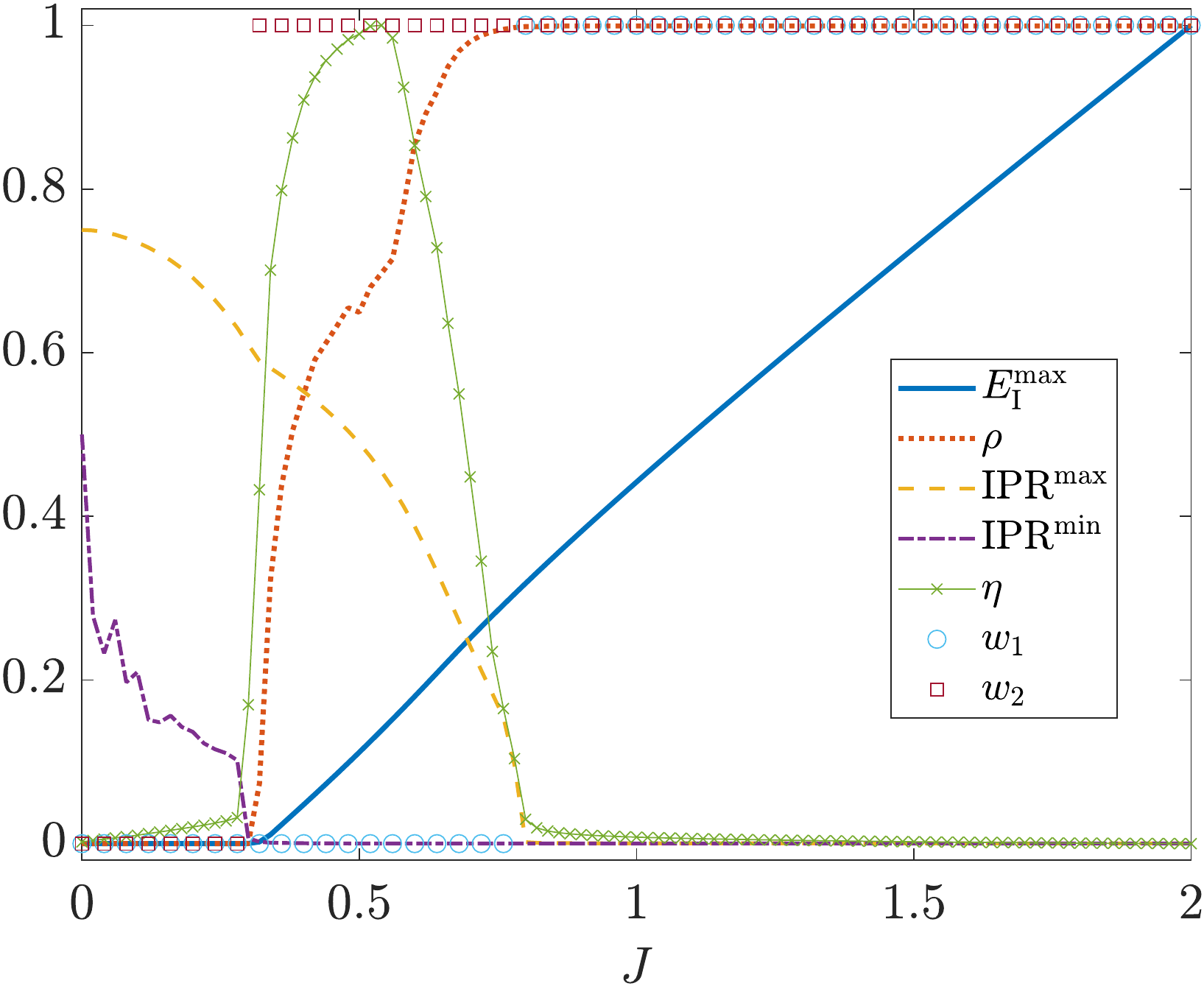}
		\par\end{centering}
	\caption{Realness of the spectrum, IPRs and winding numbers of the Model 1
		versus the unidirectional hopping amplitude $J$ \cite{Note1}. Other
		system parameters are chosen as $V=1$, $\phi=\pi/10$ and $\alpha=(\sqrt{5}-1)/2$.
		The length of lattice is $L=2584$. \label{fig:M1-E-IPR-W}}
\end{figure}

In Fig.~\ref{fig:M1-E-IPR-W}, we present the maximal imaginary part
of eigenenergies {[}Eq.~(\ref{eq:EImax}){]}, the density of states
with complex eigenvalues {[}Eq.~(\ref{eq:DOS}){]}, the maximal 
{[}Eq.~(\ref{eq:IPRmax}){]} and minimal {[}Eq.~(\ref{eq:IPRmin}){]} values
of IPRs, the smoking gun function of critical phase {[}Eq.~(\ref{eq:ETA}){]}
and the winding numbers {[}Eq.~(\ref{eq:w12}){]} of Model 1 versus
the hopping amplitude $J$ for a typical case. We observe that when
$J$ is small ($J<J_{c1}\simeq0.3$ in Fig.~\ref{fig:M1-E-IPR-W}),
all the eigenstates are indeed localized (${\rm IPR}^{\min}>0$) and their
energies are real ($E_{{\rm I}}^{\max}=\rho=0$), implying that that
system is in a ${\cal PT}$-invariant localized phase in this region.
When $J$ is large enough ($J>J_{c2}\simeq0.8$ in Fig.~\ref{fig:M1-E-IPR-W}),
all the eigenstates become extended (${\rm IPR}^{\max}\rightarrow0$)
with complex eigenvalues ($\rho\rightarrow1$), which means that the
system is in a ${\cal PT}$-broken extended phase in the large-$J$
region (we assume $V=1$ as the unit of energy). With the increase
of $J$ from zero to a finite value $J_{c1}$, the system first
undergoes a ${\cal PT}$-breaking transition signified by the emergence
of complex eigenvalues ($E_{{\rm I}}^{\max},\rho>0$). This is accompanied
by a delocalization transition through which certain eigenstates become
extended (${\rm IPR}^{\min}\rightarrow0$). The ${\cal PT}$ and delocalization
transitions at $J_{c1}$ are correctly captured by the quantized jump
of winding number $w_{2}$, which demonstrates their topological nature.
Notably, not all eigenstates become extended following the first transition
at $J_{c1}$, after which we still have localized states (${\rm IPR}^{\max}>0$).
Therefore, a critical phase in which extended and localized eigenstates
coexist appear in the intermediate region ($J_{c1}<J<J_{c2}$). States
with real and complex eigenenergies are also coexistent in this phase
($0<\rho<1$). When $J$ further increases up to $J_{c2}$, we encounter
a second transition, through which all the eigenstates become extended
and the spectrum is mostly complex. This transition is further characterized by
the quantized jump of winding number $w_{1}$. Therefore,
the collaboration between unidirectional hopping and non-Abelian quasiperiodic
potential could create a localized phase with real spectrum, an extended
phase with complex spectrum and a critical mobility edge phase with
mixed spectrum, which are separated by a ${\cal PT}$-breaking transition and
two localization transitions in the Model 1. The absence of critical
phases in the Abelian counterpart of Model 1 \cite{NHQC12} confirms
the importance of non-Abelian effects in our system.

\begin{figure}
	\begin{centering}
		\includegraphics[scale=0.5]{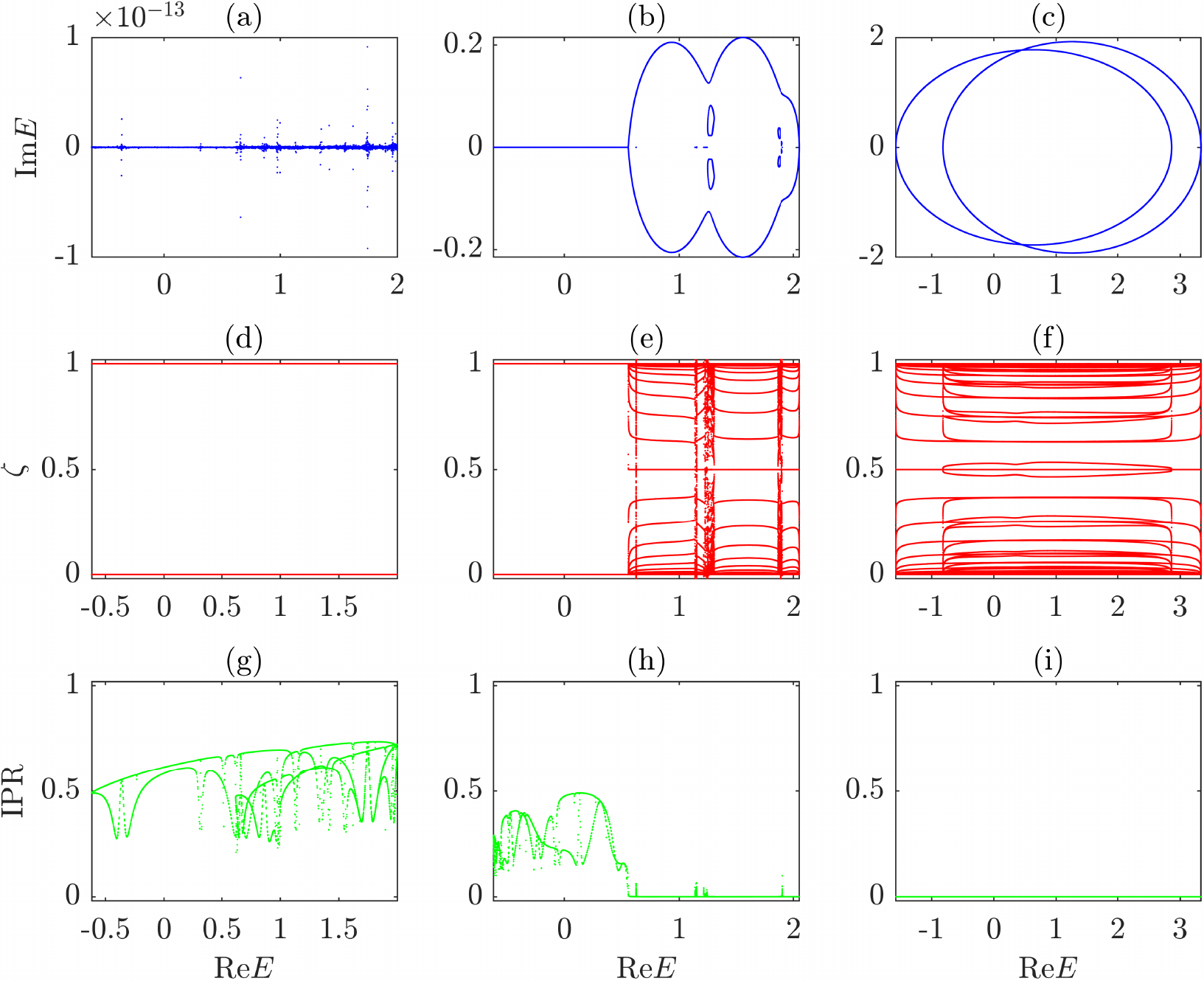}
		\par\end{centering}
	\caption{Examples of eigenenergies, ES and IPRs for different phases of the
		Model 1. The hopping amplitude is set to $J=0.1$ for (a),
		(d), (g); $J=0.5$ for (b), (e), (h); and $J=2$ for
		(c), (f), (i). Other system parameters are chosen as $V=1$, $\phi=\pi/10$
		and $\alpha=(\sqrt{5}-1)/2$ for all panels. The length of lattice
		is $L=2584$. \label{fig:M1-E-ES-IPR}}
\end{figure}

To gain a deeper understanding about the mobility edges of the spectrum
in the critical phase, we show the eigenenergies on the complex plane,
the ES and the IPRs versus the real part of energy for three typical
cases of the Model 1 in Fig.~\ref{fig:M1-E-ES-IPR}. At each ${\rm Re}E$,
the ES is obtained by first filling all states whose real parts of
energies are below ${\rm Re}E$, and then following the procedure outlined
in Sec.~\ref{sec:MS}. We see that when $J<J_{c1}$, all the eigenstates
indeed have finite IPRs. The ES is pinned around $\zeta=0$ and
$1$, suggesting vanishing contributions to the EE according to Eq.~(\ref{eq:EE}).
This is expected, as each bulk state in this case is localized around
a certain unit cell in either the subsystem A or B, not
both. No signatures of mobility edges are observed in the spectrum.
When $J>J_{c2}$, all the eigenstates are extended with vanishing
IPRs, and a large portion of ES deviates from $\zeta=0,1$. The energy spectrum
form two loops on the complex plane and no mobility edges are observed.
The most interesting situation appears when $J=0.5\in(J_{c1},J_{c2})$.
From the IPRs in Fig.~\ref{fig:M1-E-ES-IPR}(h), we find localized
states not only below a certain ${\rm Re}E$ but also at higher energies.
The eigenenergies of these localized states are all real, as observed
in Fig.~\ref{fig:M1-E-ES-IPR}(b). Therefore, the system could possess
multiple mobility edges at different ${\rm Re}E$ in the critical
phase. These mobility edges show clear signatures in the ES of 
Fig.~\ref{fig:M1-E-ES-IPR}(e). In Abelian NHQCs, signatures of mobility
edges have been found in the ES \cite{NHQC39}. Our results demonstrate
that more complicated structures of mobility edges in non-Abelian
NHQCs can also be identified from the distribution of ES at different
energies.

\begin{figure}
	\begin{centering}
		\includegraphics[scale=0.47]{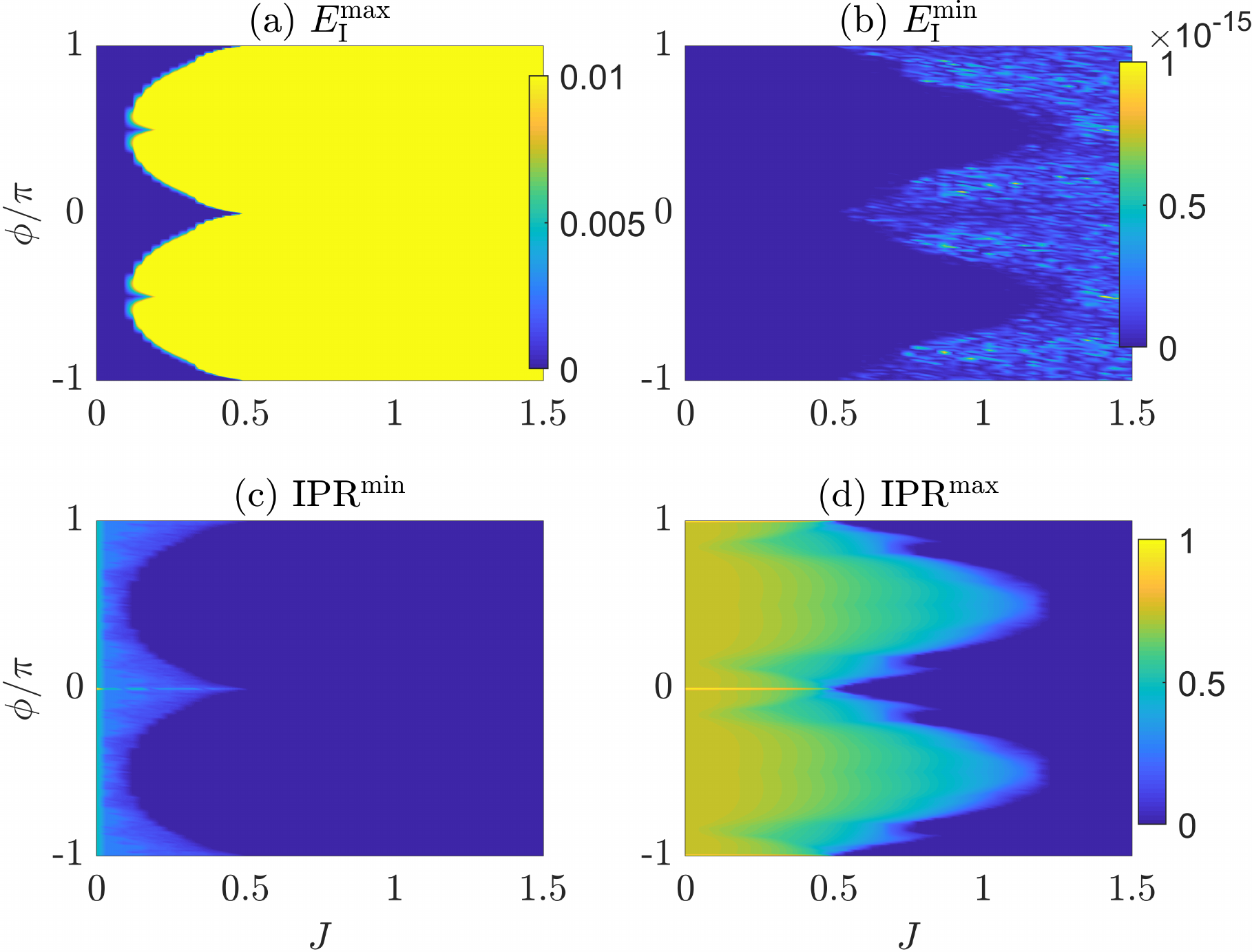}
		\par\end{centering}
	\caption{The maximum and minimum of the imaginary parts of eigenenergies and
		the IPRs versus the hopping amplitude $J$ and the Abelian phase $\phi$
		for the Model 1. Other system parameters are $V=1$ and $\alpha=(\sqrt{5}-1)/2$
		for all panels. The length of lattice is $L=610$. The panels (c)
		and (d) share the same color bar. \label{fig:M1-MaxMin-E-IPR}}
\end{figure}

To check the spectral and localization transitions in more general
situations, we present the extreme values of ${\rm Im}E$ 
{[}Eqs.~(\ref{eq:EImax}) and (\ref{eq:EImin}){]} and IPRs {[}Eqs.~(\ref{eq:IPRmax})
and (\ref{eq:IPRmin}){]} versus the hopping amplitude $J$ and the
Abelian phase factor $\phi$ in Fig.~\ref{fig:M1-MaxMin-E-IPR}. We
find that the ${\cal PT}$-transition point could vary with $\phi$
in a non-monotonic manner. Yet it is coincident with the boundary
between extended and critical phases at every $\phi$, as shown in
Figs.~\ref{fig:M1-MaxMin-E-IPR}(a) and \ref{fig:M1-MaxMin-E-IPR}(c).
Similarly, when the minimal values of ${\rm Im}E$ start to deviate
from zero, the last localized eigenstate with ${\rm IPR}^{\max}>0$
vanishes and all the states become extended, as shown in Figs.~\ref{fig:M1-MaxMin-E-IPR}(b)
and \ref{fig:M1-MaxMin-E-IPR}(d). Note in passing that at $\phi=0,\pm\pi$,
the system can be reduced to two equivalent copies of Abelian unidirectional
AAH models. Each of them has only a single transition point at $J_{c}=V/2$
and holding no critical phases \cite{NHQC12}. Therefore, the presence
of extended, critical, localized phases and the transitions among
them are robust to the variation of phase factor $\phi$ ($\neq0,\pm\pi$)
in our non-Abelian NHQC Model 1.

\begin{figure}
	\begin{centering}
		\includegraphics[scale=0.47]{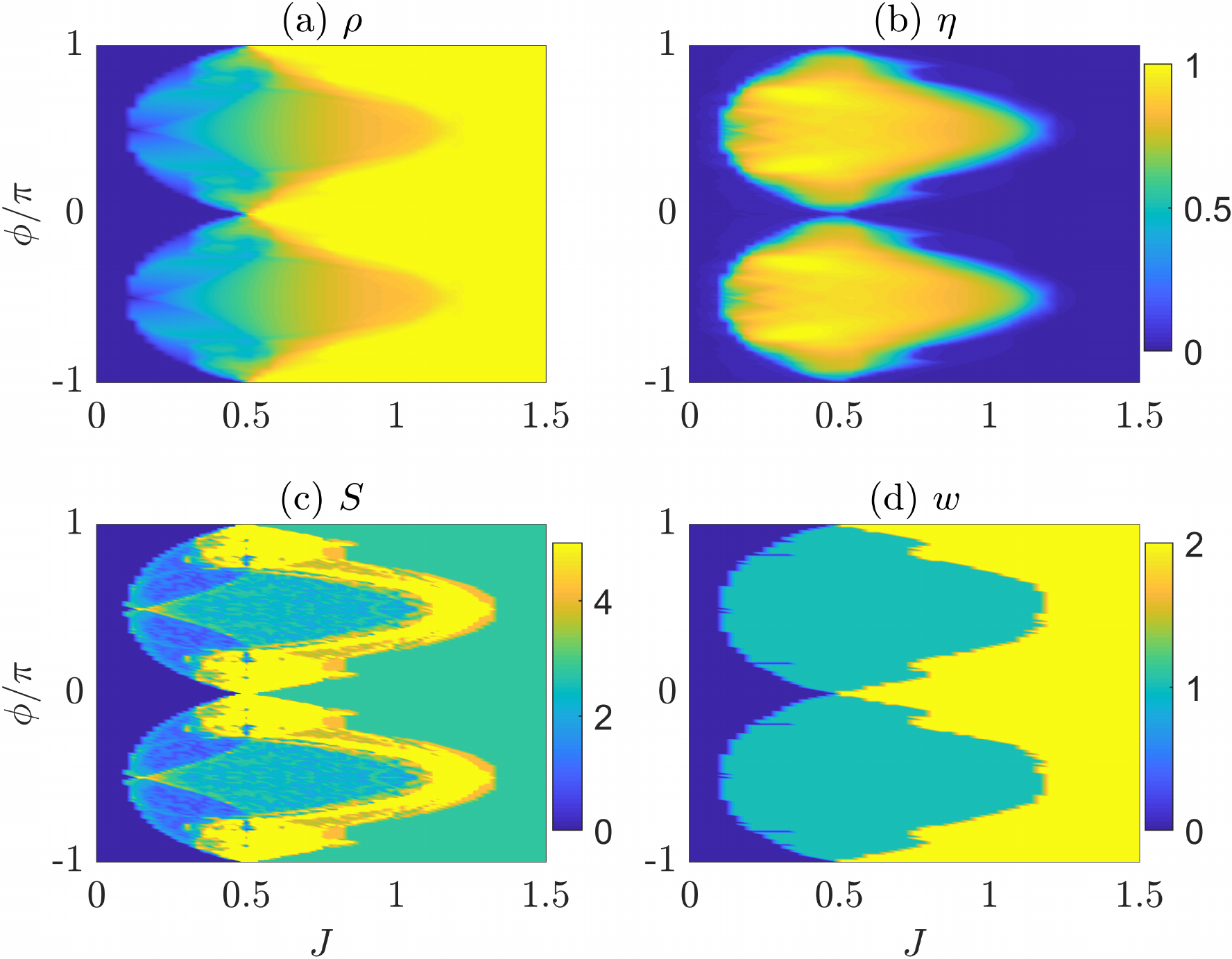}
		\par\end{centering}
	\caption{The density of states with real energies, the smoking gun function
		of critical phase, the EE and the winding numbers versus the hopping
		amplitude $J$ and the Abelian phase $\phi$ for the Model 1 \cite{Note1}.
		Other system parameters are $V=1$ and $\alpha=(\sqrt{5}-1)/2$ for
		all panels. The length of lattice is $L=610$. The panels (a) and
		(b) share the same color bar. \label{fig:M1-DOS-ETA-EE-W}}
\end{figure}

Finally, we establish the phase diagram of our Model 1 from the entanglement
and topological perspectives. In Fig.~\ref{fig:M1-DOS-ETA-EE-W},
we show the density of states with complex eigenvalues {[}Eq.~(\ref{eq:DOS}){]},
the smoking gun function of critical phase {[}Eq.~(\ref{eq:ETA}){]},
the EE {[}Eq.~(\ref{eq:EE}){]} and the phase diagram determined by
the winding number {[}Eq.~(\ref{eq:w12}){]} of the Model 1. Compared
with the results of IPRs in Fig.~\ref{fig:M1-MaxMin-E-IPR}, we clearly
see that $\rho\rightarrow0$ in the ${\cal PT}$-invariant localized
phase, $\rho\rightarrow1$ in the ${\cal PT}$-broken extended phase,
and $0<\rho<1$ in the critical mobility edge phase highlighted by
the $\eta$ in Fig.~\ref{fig:M1-DOS-ETA-EE-W}(b). The EE is obtained
by filling all the eigenstates with real energies at each given $(J,\phi)$
and following the steps in Sec.~\ref{sec:MS}. In Fig.~\ref{fig:M1-DOS-ETA-EE-W}(c),
the EE is found to vanish throughout the localized phase, fluctuating
in the critical phase while keeping a constant value $S\simeq4\ln2$
in the extended phase. It can thus be employed as a good entanglement-based
character to distinguish different phases in non-Abelian NHQCs. In
Fig.~\ref{fig:M1-DOS-ETA-EE-W}(d), the blue, green and yellow regions
refer to the localized, critical and extended phases, respectively.
The boundary between the blue (green) and green (yellow) regions is
the boundary around which the winding number $w_{2}$ ($w_{1}$) takes
a quantized jump. These boundaries are well consistent with the phase
boundaries identified from $\rho$, IPRs and $S$. Therefore, the winding
numbers $(w_{1},w_{2})$ serve as topological order parameters to
characterize the ${\cal PT}$ and localization transitions in our
system. Putting together, we conclude that the cooperation between
unidirectional hoppings and non-Abelian effects could indeed create
many intriguing phases and transitions in a quasiperiodic lattice.
In the following subsections, we will demonstrate that this physical
picture holds for more general types of non-Hermitian effects and
non-Abelian potentials.

\subsection{Nonreciprocal hopping\label{subsec:M2}}

We now consider a non-Abelian NHAAH model with nonreciprocal hopping
amplitudes, which can be viewed as a generalized version of the Model
1. According to the Eqs.~(\ref{eq:Seq}), (\ref{eq:Theta}) and Table
\ref{tab:Mod}, the eigenvalue equation of this Model 2 in position
representation takes the form
\begin{equation}
	Je^{-\beta}\boldsymbol{\psi}_{n+1}+Je^{\beta}\boldsymbol{\psi}_{n-1}+V(e^{-i\phi}\Theta_{n}+e^{i\phi}\Theta_{n}^{-1})\boldsymbol{\psi}_{n}=E\boldsymbol{\psi}_{n},\label{eq:SeqM2}
\end{equation}
where $\Theta_{n}=e^{i2\pi\alpha n\sigma_{y}}e^{i2\pi\alpha n\sigma_{z}}$.
The non-Hermitian effect is now introduced by an imaginary phase factor
$i\beta$ accompanying the nearest-neighbor hopping amplitude. In
the Abelian counterpart of this model, a ${\cal PT}$-transition together
with a localization transition is predicted at $\beta_{c}=\ln[V/(2J)]$
for any irrational $\alpha$, which is further characterized by the
quantized jump of a spectral winding number from zero to one \cite{NHQC3}.
When $|\beta|<|\beta_{c}|$ ($|\beta|>|\beta_{c}|$), all the eigenstates
of the Abelian NHQC are localized (extended) with a real (complex)
spectrum under the PBC \cite{NHQC3}. The non-Abelian Model 2 also
possesses the ${\cal PT}$ symmetry, albeit different from its Abelian
cousin in its explicit form as discussed in Sec.~\ref{sec:MM}. Therefore,
we expect the system to reside in a localized (an extended) phase
with real (complex) spectrum in the limit $\beta\rightarrow0$ ($\beta\rightarrow\infty$)
for any irrational $\alpha$ assuming $|J|\ll|V|$. ${\cal PT}$-breaking
and localization transitions should happen at some finite values of
$\beta$ between these two limits. Furthermore, the non-Abelian potential
may also help to expand the Abelian transition point $\beta_{c}$
to a critical region $\beta\in(\beta_{c1},\beta_{c2})$, in which extended
and localized eigenstates coexist. This is similar to the situation encountered
in the Model 1. The presence of such a critical mobility edge phase
and its properties in the non-Abelian NHQC Model 2 will be uncovered
in the following discussions.

\begin{figure}
	\begin{centering}
		\includegraphics[scale=0.5]{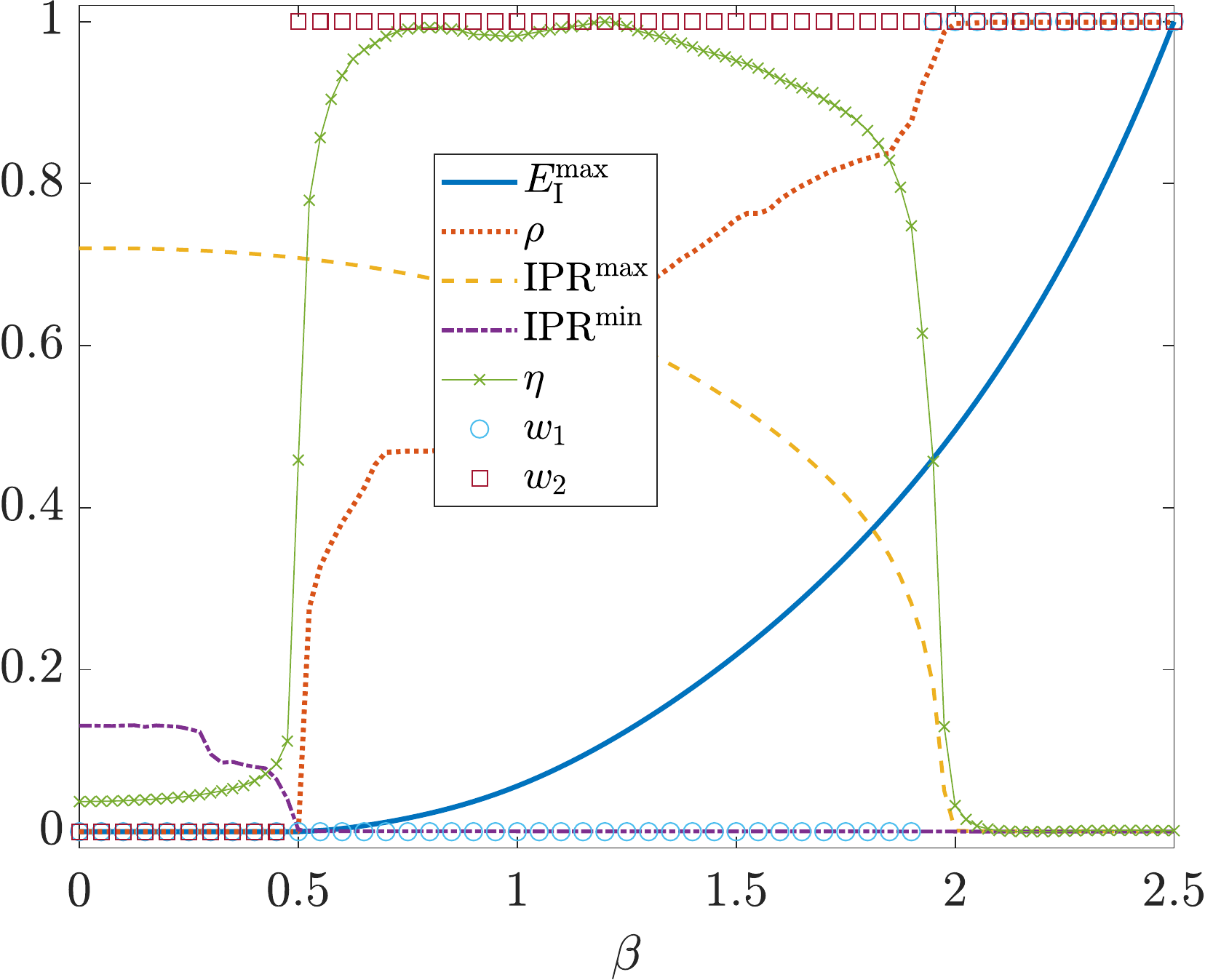}
		\par\end{centering}
	\caption{Realness of the spectrum, IPRs and winding numbers of the Model 2
		versus the nonreciprocal hopping modulation $\beta$ \cite{Note1}.
		Other system parameters are set as $J=1$, $V=6$, $\phi=\pi/2$ and
		$\alpha=(\sqrt{5}-1)/2$. The length of lattice is $L=2584$. \label{fig:M2-E-IPR-W}}
\end{figure}

In Fig.~\ref{fig:M2-E-IPR-W}, we present the maximal imaginary parts
of eigenenergies {[}Eq.~(\ref{eq:EImax}){]}, the density of states
with complex eigenvalues {[}Eq.~(\ref{eq:DOS}){]}, the maximal 
{[}Eq.~(\ref{eq:IPRmax}){]} and minimal {[}Eq.~(\ref{eq:IPRmin}){]} values
of IPRs, the smoking gun function of critical phases {[}Eq.~(\ref{eq:ETA}){]}
and the winding numbers {[}Eq.~(\ref{eq:w12}){]} of the Model 2 versus
the nonreciprocal hopping parameter $\beta$ for a typical case. From
the ${\rm IPR}^{\min}$ and ${\rm IPR}^{\max}$, we identify two transition
points at $\beta_{c1}\simeq0.5$ and $\beta_{c2}\simeq2.0$. When
$\beta<\beta_{c1}$, all the eigenstates are localized (${\rm IPR}^{\min}>0$)
and the spectrum is real ($E_{{\rm I}}^{\max},\rho=0$). The system
thus resides in a ${\cal PT}$-invariant localized phase in this small-$\beta$
region. When $\beta>\beta_{c2}$, all the eigenstates are extended
(${\rm IPR}^{\max}\rightarrow0$) and the spectrum is complex. The
system thus belongs to a ${\cal PT}$-broken extended phase in this
large-$\beta$ region. Interestingly, in the intermediate region with
$\beta_{c1}<\beta<\beta_{c2}$, the spectrum is formed by comparable
numbers of real and complex eigenvalues ($0<\rho<1$). Extended and
localized eigenstates also coexist in this region (${\rm IPR}^{\min}\rightarrow0$
yet ${\rm IPR}^{\max}>0$), yielding a critical mobility
edge phase as signaled by the function $\eta$. This critical
phase is originated from the coexistence of non-Hermitian and non-Abelian
potentials in our system. It is absent in the Abelian limit of our
Model 2 \cite{NHQC3}. Therefore, with the increase of $\beta$, the
system first undergoes a ${\cal PT}$-breaking and delocalization
transition at $\beta_{c1}$, which is also characterized by the quantized
change of winding number $w_{2}$. When $\beta$ further increases,
we encounter a second delocalization transition at $\beta_{c2}$,
which is accompanied by the quantized jump of another winding number
$w_{1}$. The two separate transitions and the emerging critical
phase are all unique to our non-Abelian NHQC model. The two transition points will merge
at $\beta_{c}=\ln[V/(2J)]\simeq1.1$ and the critical phase will disappear in the associated
Abelian model \cite{NHQC3}. These observations again demonstrate that
non-Abelian potentials could induce richer phases and transition patterns
in non-Hermitian quasiperiodic lattices.

\begin{figure}
	\begin{centering}
		\includegraphics[scale=0.5]{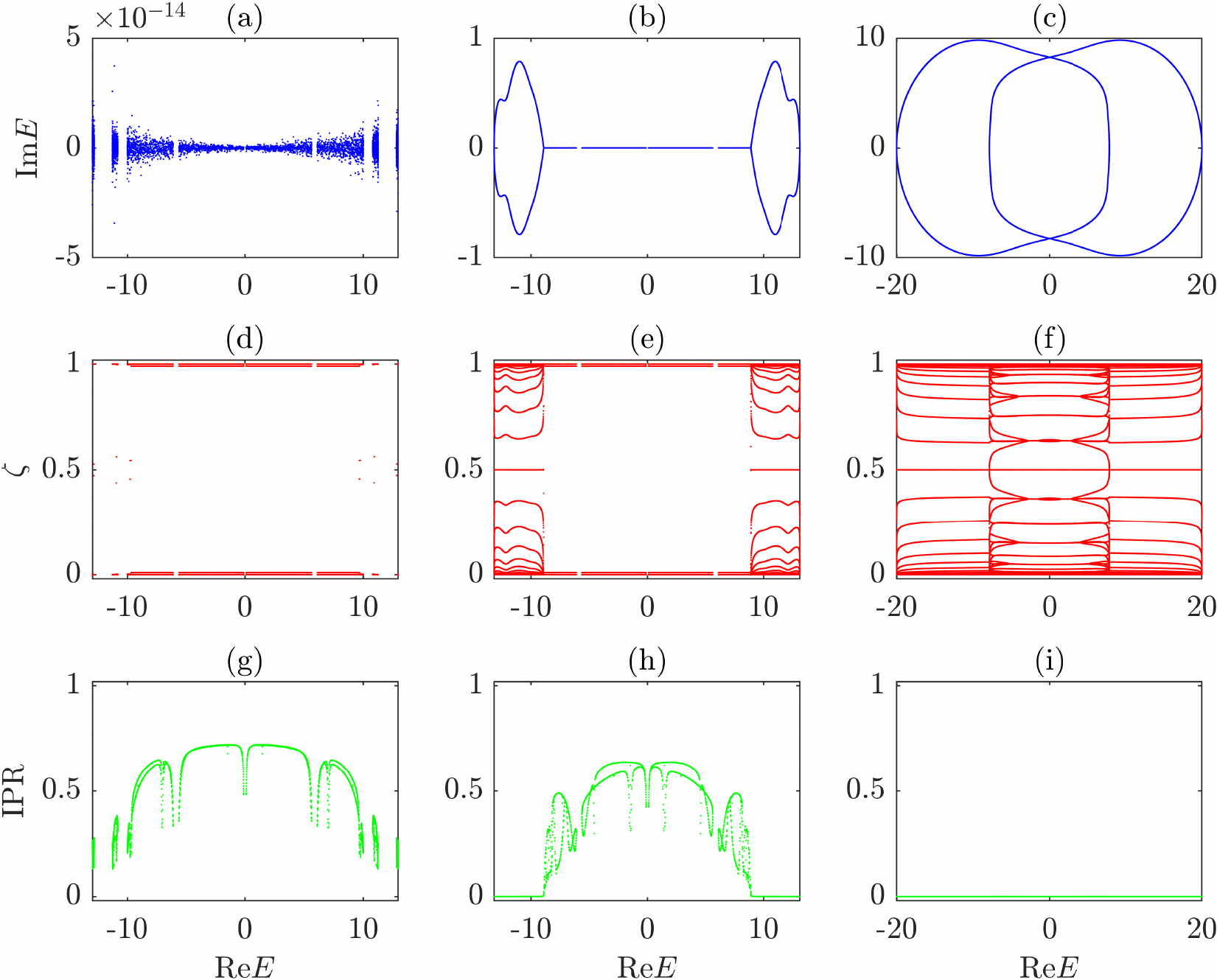}
		\par\end{centering}
	\centering{}\caption{Examples of eigenenergies, ES and IPRs for different phases of the
		Model 2. The nonreciprocal hopping modulation is set to $\beta=0.1$
		for (a), (d), (g); $\beta=1.1$ for (b), (e), (h); and
		$\beta=2.5$ for (c), (f), (i). Other system parameters are
		chosen as $J=1$, $V=6$, $\phi=\pi/2$ and $\alpha=(\sqrt{5}-1)/2$
		for all panels. The length of lattice is $L=2584$. \label{fig:M2-E-ES-IPR}}
\end{figure}

To further decode the structure of mobility edges in the critical
phase, we show the eigenenergies on the complex plane, the ES and
the IPRs versus the real parts of energies for three typical cases of
the Model 2 in Fig.~\ref{fig:M2-E-ES-IPR}. At each given ${\rm Re}E$,
the ES is obtained by first filling all the eigenstates whose real
parts of energies are below ${\rm Re}E$, and then following the procedure
outlined in Sec.~\ref{sec:MM}. We find that for the case with $\beta<\beta_{c1}$,
all the eigenstates are indeed localized with finite IPRs and real
energies. The ES is mostly pinned around $\zeta=0,1$ and no signatures
of mobility edges are observed. A few ES values that deviate sufficiently
away from $\zeta=0,1$ in Fig.~\ref{fig:M2-E-ES-IPR}(d) may originate
from eigenmodes that are localized around the entanglement cuts between
the subsystems A and B. In the case with $\beta>\beta_{c2}$, all
the eigenmodes are found to be extended with vanishing IPRs. The spectrum
form two loops on the complex plane in Fig.~\ref{fig:M2-E-ES-IPR}(c)
and the eigenvalues are mostly complex. The ES contains notable portions
that are away from $\zeta=0,1$ at every ${\rm Re}E$, and no
signatures of mobility edges are observed. In the intermediate case
with $\beta=1.1\in(\beta_{c1},\beta_{c2})$, we observe mobility edges
at two different energies $E_{\pm}$ that are symmetric along the
${\rm Re}E$ axis, as shown in Figs.~\ref{fig:M2-E-ES-IPR}(b) and
\ref{fig:M2-E-ES-IPR}(h). Eigenstates whose ${\rm Re}E\in(E_{-},E_{+})$
are localized with real energies, and otherwise extended with complex
energies. These mobility edges can be clearly identified from the
ES as shown in Fig.~\ref{fig:M2-E-ES-IPR}(e). Therefore, we can also
use the ES as an information-based detector to find mobility
edges in the spectrum and separate localized from delocalized states
in our non-Abelian NHQC Model 2.

\begin{figure}
	\begin{centering}
		\includegraphics[scale=0.47]{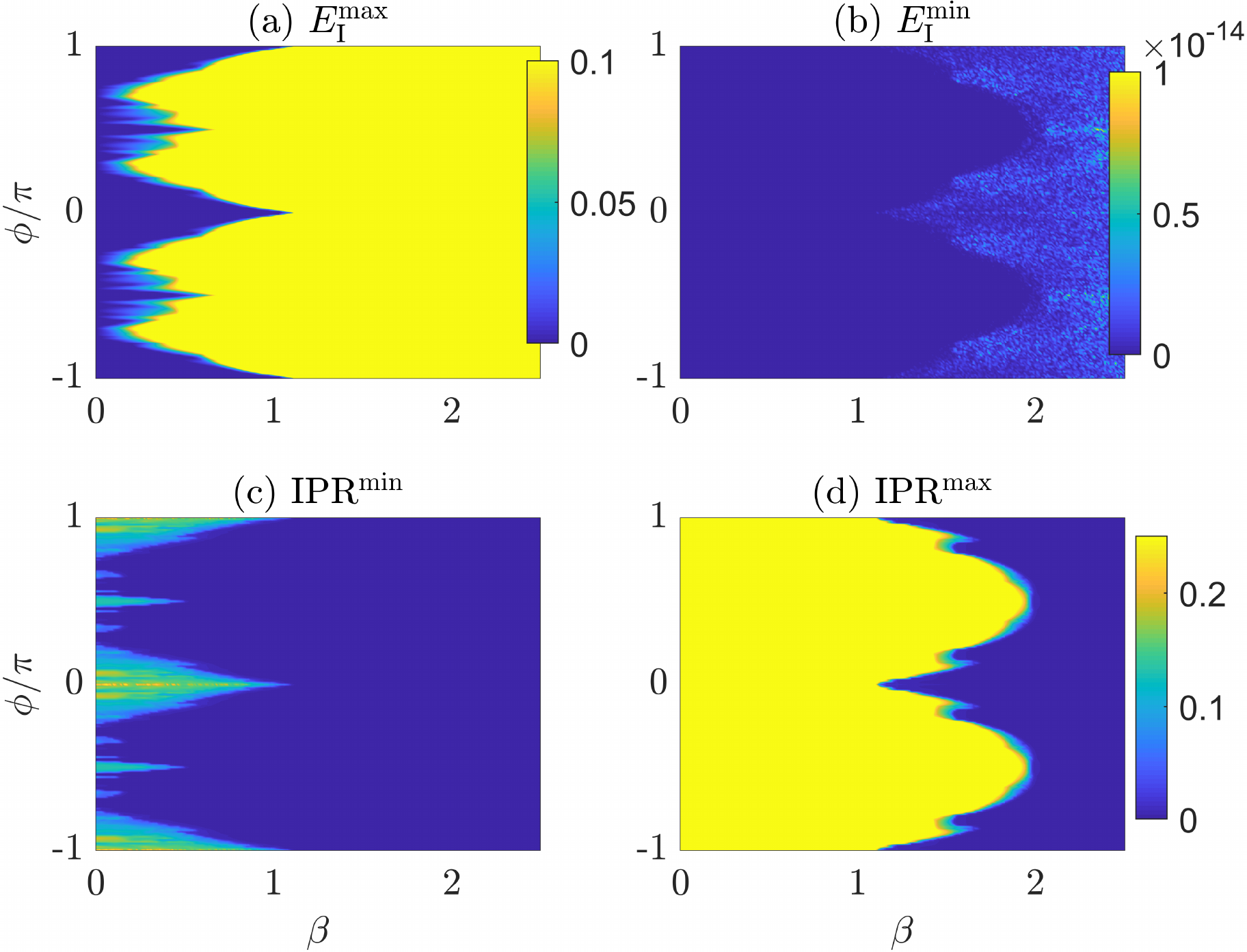}
		\par\end{centering}
	\caption{The maximum and minimum of the imaginary parts of eigenenergies and
		the IPRs versus the nonreciprocal hopping modulation $\beta$ and
		the Abelian phase $\phi$ for the Model 2. Other system parameters
		are $J=1$, $V=6$ and $\alpha=(\sqrt{5}-1)/2$ for all panels. The
		length of lattice is $L=610$. The panels (c) and (d) share the same
		color bar. \label{fig:M2-MaxMin-E-IPR}}
\end{figure}

To check the ${\cal PT}$-breaking and localization transitions of
our Model 2 in more general situations, we show the extreme values
of ${\rm Im}E$ {[}Eqs.~(\ref{eq:EImax}) and (\ref{eq:EImin}){]}
and IPRs {[}Eqs.~(\ref{eq:IPRmax}) and (\ref{eq:IPRmin}){]} versus
the nonreciprocal hopping parameter $\beta$ and the Abelian phase
factor $\phi$ in Fig.~\ref{fig:M2-MaxMin-E-IPR}. We observe that
when the $E_{{\rm I}}^{\max}$ starts to deviate from zero, the ${\rm IPR}^{\min}$
goes to zero, which means that the ${\cal PT}$-breaking transition
and the transition between localized and critical phases go hand-in-hand
with each other in our system. This is true for every $\phi\in(-\pi,0)\cup(0,\pi)$.
At $\phi=0,\pm\pi$, the critical phase vanishes due to the reducibility
of our Model 2 to two identical copies of Abelian NHQCs \cite{NHQC3},
as mentioned before. Moreover, the boundary where $E_{{\rm I}}^{\min}$
starts to deviate from zero is coincident with the boundary across which
the ${\rm IPR}^{\max}$ goes to zero, which means that most eigenstates
have complex energies after the transition from critical to localized
phases. This second phase boundary has a shape that can vary with
$\phi$ in a non-monotonous manner, and it merges with the first phase
boundary at $\beta_{c}=\ln[J/(2V)]$ when $\phi=0,\pm\pi$. Therefore,
we conclude that our non-Abelian NHQC Model 2 indeed holds extended,
critical and localized phases. It could further transform among them
at different values of $\beta$ and $\phi$ due to the interplay between
nonreciprocal and non-Abelian effects.

\begin{figure}
	\begin{centering}
		\includegraphics[scale=0.47]{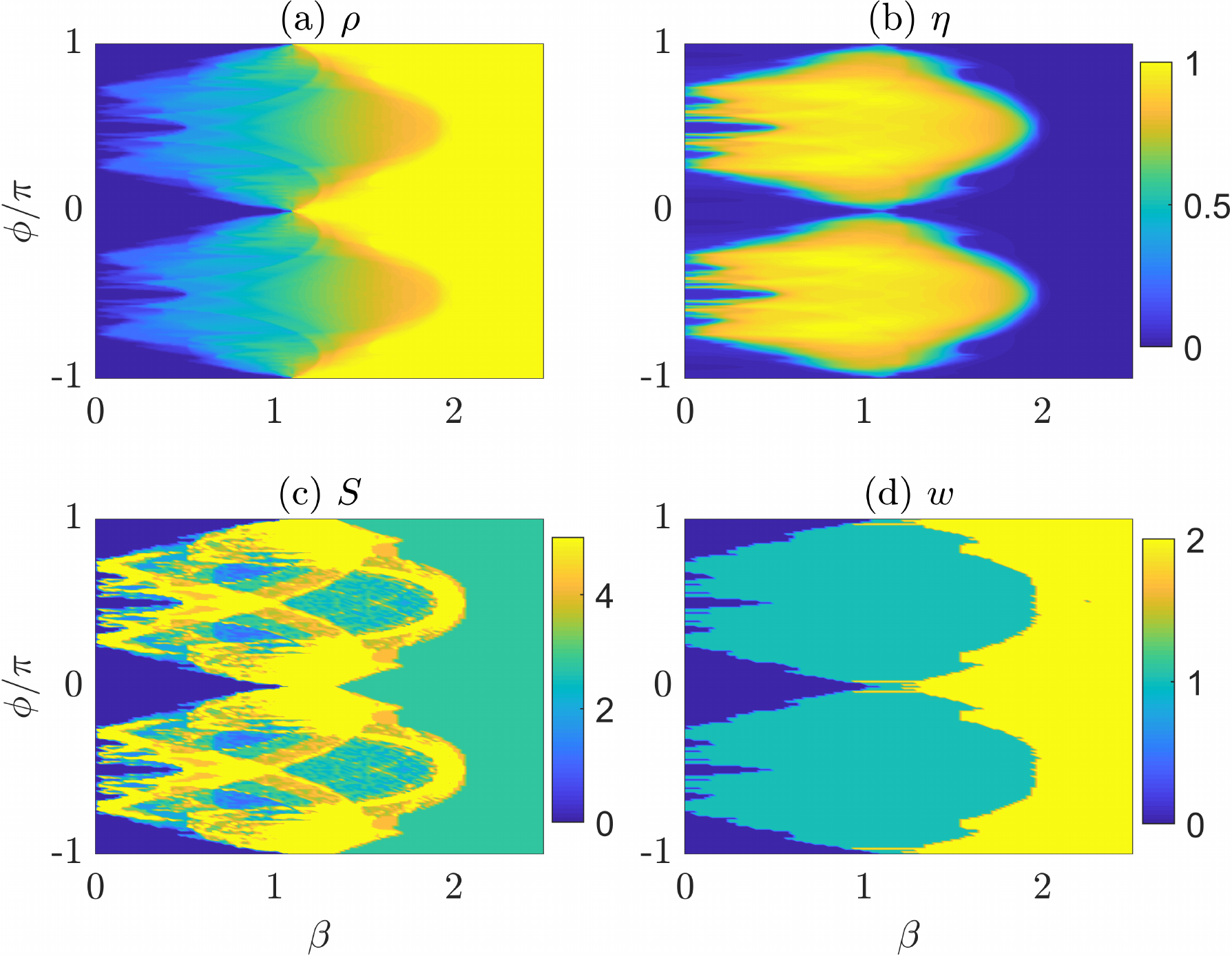}
		\par\end{centering}
	\caption{The density of states with real energies, the smoking gun function
		of critical phase, the EE and the winding numbers versus the nonreciprocal
		hopping modulation $\beta$ and the Abelian phase factor $\phi$ for
		the Model 2 \cite{Note1}. Other system parameters are $J=1$, $V=6$
		and $\alpha=(\sqrt{5}-1)/2$ for all panels. The length of lattice
		is $L=610$. The panels (a) and (b) share the same color bar. \label{fig:M2-DOS-ETA-EE-W}}
\end{figure}

We next build the phase diagram of our Model 2 from its entanglement
and topological features. In Fig.~\ref{fig:M2-DOS-ETA-EE-W}, we show
the density of states with complex eigenvalues {[}Eq.~(\ref{eq:DOS}){]},
the smoking-gun function of critical phases {[}Eq.~(\ref{eq:ETA}){]},
the EE {[}Eq.~(\ref{eq:EE}){]} and the phase diagram decided by the
winding numbers {[}Eq.~(\ref{eq:w12}){]} of Model 2. Comparing the
Fig.~\ref{fig:M2-DOS-ETA-EE-W}(a) with the spectrum and IPRs in
Fig.~\ref{fig:M2-MaxMin-E-IPR}, we see that throughout the considered
parameter regime, the density of states
$\rho\rightarrow0$ and $\rho\rightarrow1$ in the real-spectrum localized
phase and complex-spectrum extended phase, respectively. In the critical phase highlighted
by $\eta$ in the Fig.~\ref{fig:M2-DOS-ETA-EE-W}(b), we have $0<\rho<1$,
implying that real-energy localized states and complex-energy extended
states coexist in this region. The EE in Fig.~\ref{fig:M2-DOS-ETA-EE-W}(c)
is obtained by filling all the eigenstates with real energies at each
$(\beta,\phi)$ and following the recipe discussed in Sec.~\ref{sec:MS}.
We find the same EE $S=0$ ($S\simeq4\ln2$) in the ${\cal PT}$-invariant
(${\cal PT}$-broken) localized (extended) phase, and a fluctuating
$S$ in the critical mobility edge phase. Therefore, we can use the
EE to clearly distinguish phases with different localization nature
in our non-Abelian NHQC Model 2. Finally, by investigating the boundaries
where the winding numbers $w_{2}$ and $w_{1}$ get quantized jumps,
we obtain the borders of localized-to-critical and critical-to-extended
phase transitions, which are well consistent with the boundaries predicted
by the $\rho$, ${\rm IPR}^{\max,\min}$ and $S$. The winding numbers
$w_{1,2}$ can thus be adopted as topological order parameters to
characterize the ${\cal PT}$ and localization transitions in our
system. Fig.~\ref{fig:M2-DOS-ETA-EE-W}(d) yields the phase diagram
of Model 2, in which the blue, green and yellow regions correspond
to the localized, critical and extended phases, respectively. To sum
up, we find that the combined efforts of nonreciprocal hoppings and
non-Abelian effects could also generate multiple intriguing phases
with different localization properties and rich phase transitions
in a quasiperiodic system. In the next subsection, we further
explore the effects of a non-Abelian non-Hermitian onsite potential
in generating these new phases.

\subsection{Complex onsite potential\label{subsec:M3}}

In the last example, we consider an NHAAH model with onsite gain and
loss in a non-Abelian quasiperiodic potential. Following the Eqs.~(\ref{eq:Theta}),
(\ref{eq:Seq}) and Table \ref{tab:Mod}, the eigenvalue equation
of this Model 3 in the lattice representation reads
\begin{equation}
	J\boldsymbol{\psi}_{n+1}+J\boldsymbol{\psi}_{n-1}+V(e^{-i\phi}\Theta_{n}+e^{i\phi}\Theta_{n}^{-1})\boldsymbol{\psi}_{n}=E\boldsymbol{\psi}_{n},\label{eq:SeqM3}
\end{equation}
where $\Theta_{n}=e^{i(2\pi\alpha n+i\gamma)\sigma_{y}}e^{i(2\pi\alpha n+i\gamma)\sigma_{z}}$.
The hopping amplitude $J$ is now symmetric and the non-Hermitian
effect is introduced solely by the imaginary non-Abelian phase factor
$i\gamma$. In the Abelian counterpart of the model, a ${\cal PT}$-breaking
and localization transition can happen at $\gamma_{c}=\ln(2J/V)$
under the PBC for any irrational $\alpha$, which is signified by
the unit jump of a spectral winding number \cite{NHQC4}. All the
eigenstates are extended (localized) with real (complex) eigenvalues
when $|\gamma|<|\gamma_{c}|$ ($|\gamma|>|\gamma_{c}|$), and no critical
mobility edge phases are identified in the Abelian model \cite{NHQC4}.
Our Model 3 can be reduced to two equivalent copies of this Abelian
model when we take $\phi=0,\pm\pi$, with a critical point
at $\gamma'_{c}=(1/2)\ln(2J/V)$. Since the Model 3 also possesses
the ${\cal PT}$-symmetry as discussed in Sec.~\ref{sec:MM}, we expect
the system to exhibit an extended (a localized) phase with a real
(complex) spectrum in the limit $\gamma\rightarrow0$ ($\gamma\rightarrow\infty$)
for any irrational $\alpha$ assuming $|J|\ll|V|$. ${\cal PT}$-breaking
and localization transitions must happen between these two limits
with the increase of $\gamma$. However, the non-Abelian potential
may again extend the critical point $\gamma'_{c}$ into a critical
regime $\gamma\in(\gamma_{c1},\gamma_{c2})$, in which extended and
localized eigenstates can coexist and are separated by mobility edges.
Eigenstates with real and complex energies may also survive together
in this critical phase. In the following, we unveil the existence
of such a critical phase and characterize the transitions induced
by non-Abelian non-Hermitian potentials between different phases in
our Model 3.

\begin{figure}
	\begin{centering}
		\includegraphics[scale=0.5]{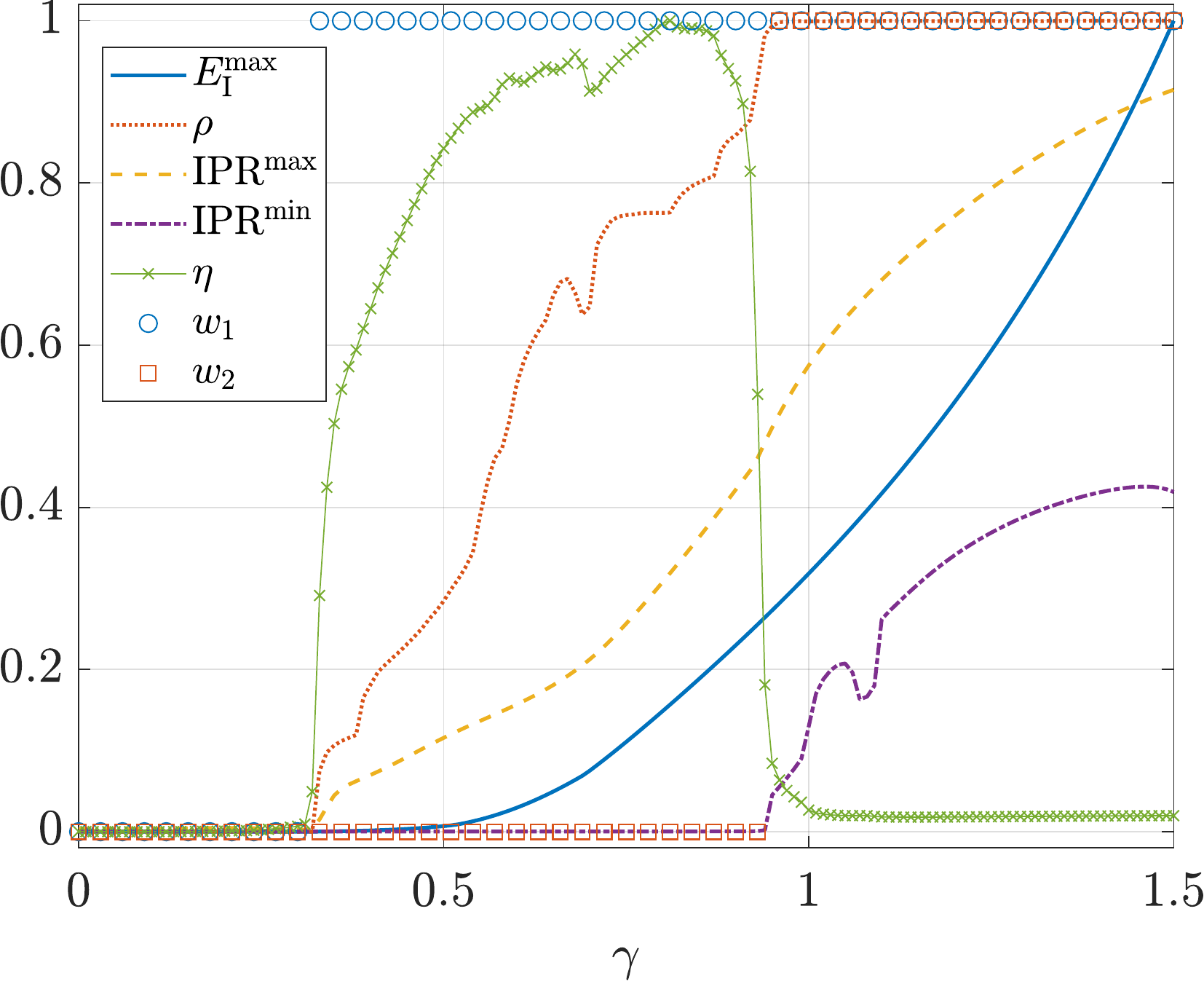}
		\par\end{centering}
	\caption{Realness of the spectrum, IPRs and winding numbers of the Model 3
		versus the imaginary non-Abelian phase $\gamma$ \cite{Note1}. Other
		system parameters are set as $J=1$, $V=0.5$, $\phi=\pi/2$ and $\alpha=(\sqrt{5}-1)/2$.
		The length of lattice is $L=2584$. \label{fig:M3-E-IPR-W}}
\end{figure}

In Fig.~\ref{fig:M3-E-IPR-W}, we present the maximal ${\rm Im}E$
of all states {[}Eq.~(\ref{eq:EImax}){]}, the density of states with
complex energies {[}Eq.~(\ref{eq:DOS}){]}, the maximal {[}Eq.~(\ref{eq:IPRmax}){]}
and minimal {[}Eq.~(\ref{eq:IPRmin}){]} IPRs of all states, the smoking-gun
 function of critical phases {[}Eq.~(\ref{eq:ETA}){]} and the winding
numbers {[}Eq.~(\ref{eq:w12}){]} of the Model 3 versus the imaginary non-Abelian
phase shift $\gamma$ for a typical example. From the IPRs, we can
identify two localization transition points $\gamma_{c1}\simeq0.31$
and $\gamma_{c2}\simeq0.94$ with the deviations of ${\rm IPR}^{\max}$
and ${\rm IPR}^{\min}$ from zero during the increase of $\gamma$,
respectively. $\gamma_{c1}$ is also consistent with a ${\cal PT}$-breaking
transition point, after which complex eigenvalues in energy start to
emerge (with $E_{{\rm I}}^{\max}>0$). When $\gamma<\gamma_{c1}$,
all the eigenstates are extended (${\rm IPR}^{\max}\rightarrow0$)
and carrying real energies ($E_{{\rm I}}^{\max},\rho=0$). The system
is thus in a ${\cal PT}$-invariant metallic phase in this region.
When $\gamma>\gamma_{c2}$, all the eigenstates are localized (${\rm IPR}^{\min}>0$)
with most of them having complex energies ($\rho\rightarrow1$). The
system is thus in a ${\cal PT}$-broken insulating phase in this regime.
When $\gamma\in(\gamma_{c1},\gamma_{c2})$, extended (${\rm IPR}^{\min}\rightarrow0$)
and localized (${\rm IPR}^{\max}>0$) eigenstates are found to coexist.
Real and complex eigenvalues occupy comparable portions in the
spectrum ($0<\rho<1$). This intermediate region thus corresponds
to a critical mobility edge phase, as clearly highlighted by the function
$\eta$ in Fig.~\ref{fig:M3-E-IPR-W}. Moreover, we observe a quantized
jump of the winding number $w_{1}$ ($w_{2}$) when $\gamma$ is swept
across the transition point between extended (critical) and critical
(localized) phases. The winding numbers $w_{1,2}$ can thus be used
as topological order parameters to characterize the two different
localization transitions in our non-Abelian NHQC Model 3. Note that
these two transition points will merge into a single one at $\gamma'_{c}=\ln(4)/2\simeq0.69$
if $\phi=0,\pm\pi$, for which our Model 3 is reduced to two Abelian
copies of the model considered in Ref.~\cite{NHQC4}. The two localization
transitions and the critical phase are thus all rooted in the presence
of a non-Abelian non-Hermitian quasiperiodic potential in the system,
through which richer patterns of phases and transitions 
beyond those in the Abelian counterpart of our Model 3 are generated
\cite{NHQC4}.

\begin{figure}
	\begin{centering}
		\includegraphics[scale=0.5]{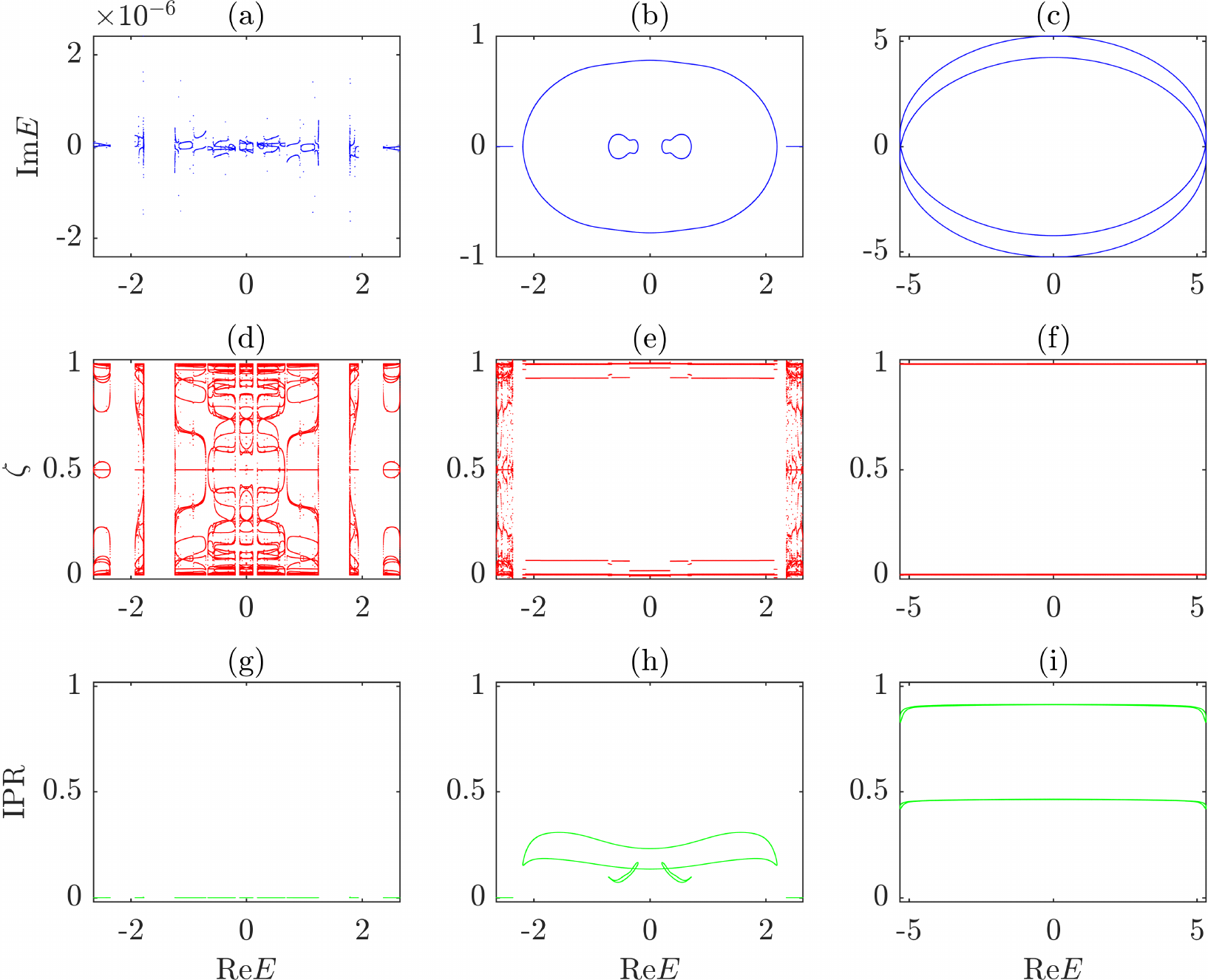}
		\par\end{centering}
	\caption{Examples of the eigenenergies, ES and IPRs for different phases of
		the Model 3. The imaginary phase is set to $\gamma=0.1$ for 
		(a), (d), (g); $\gamma=0.8$ for (b), (e), (h); and $\gamma=1.5$
		for (c), (f), (i). Other system parameters are chosen as $J=1$,
		$V=0.5$, $\phi=\pi/2$ and $\alpha=(\sqrt{5}-1)/2$ for all panels.
		The length of lattice is $L=2584$. \label{fig:M3-E-ES-IPR}}
\end{figure}

To further analyze the structure of mobility edges induced by the
non-Abelian potential, we present the spectrum on the complex plane,
the ES and the IPRs versus the real part of energy for three typical
cases of the Model 3 in Fig.~\ref{fig:M3-E-ES-IPR}. At each given
${\rm Re}E$, the ES is obtained by first filling all the eigenstates
whose real parts of energies are below ${\rm Re}E$, and then following the steps
outlined in Sec.~\ref{sec:MS}. We find that for $\gamma>\gamma_{c2}$,
all the eigenstates are indeed localized with finite IPRs, and the
spectrum is constituted by two loops on the complex plane, as shown
in Figs.~\ref{fig:M3-E-ES-IPR}(i) and \ref{fig:M3-E-ES-IPR}(c).
The ES in Fig.~\ref{fig:M3-E-ES-IPR}(f) is pinned around $\zeta=0,1$
for all ${\rm Re}E$, which implies that the states below any energy
are localized ether in the subsystem A or B of the lattice. The system
in this case then resides in a localized phase and no signatures of
mobility edges are observable. For $\gamma<\gamma_{c1}$, the spectrum
is real and the IPRs of all states are vanishingly small, as shown
in Figs.~\ref{fig:M3-E-ES-IPR}(a) and \ref{fig:M3-E-ES-IPR}(g).
From the ES in Fig.~\ref{fig:M3-E-ES-IPR}(d), we can see gaps in
certain ranges of ${\rm Re}E$. Meanwhile, sufficient parts of ES
are away from $\zeta=0,1$ in other energy regions. In this case,
the system belongs to a ${\cal PT}$-invariant extended phase with
no signals of mobility edges. When $\gamma=0.8\in(\gamma_{c1},\gamma_{c2})$,
the spectrum contains both real and complex eigenvalues, as shown
in Fig.~\ref{fig:M3-E-ES-IPR}(b). Referring to the IPRs in Fig.~\ref{fig:M3-E-ES-IPR}(h),
we realize that the extended and localized eigenmodes separately have
real and complex eigenvalues. Notably, these two types of states are
split not only by mobility gaps but also by energy gaps at different
${\rm Re}E$, which is distinct from the first two models considered
in this section. In Fig.~\ref{fig:M3-E-ES-IPR}(e), we also observe
gaps along the ${\rm Re}E$ axis. They separate states whose ES are
close to $\zeta=0,1$ from the other states whose ES are distributed
throughout the range $\zeta\in[0,1]$. The energy and mobility gaps
of the Model 3 are thus clearly identifiable from the ES. Therefore,
we can apply the ES as a detector to seek for the energy and mobility
gaps in the spectrum and distinguish the localized from delocalized
states in our non-Abelian NHQC Model 3. The results presented in 
Figs.~\ref{fig:M1-E-ES-IPR}, \ref{fig:M2-E-ES-IPR} and \ref{fig:M3-E-ES-IPR}
also revealed the generality of ES as a tool to characterize the mobility
edges in quasicrystals with different types of non-Hermitian and non-Abelian
effects.

\begin{figure}
	\begin{centering}
		\includegraphics[scale=0.47]{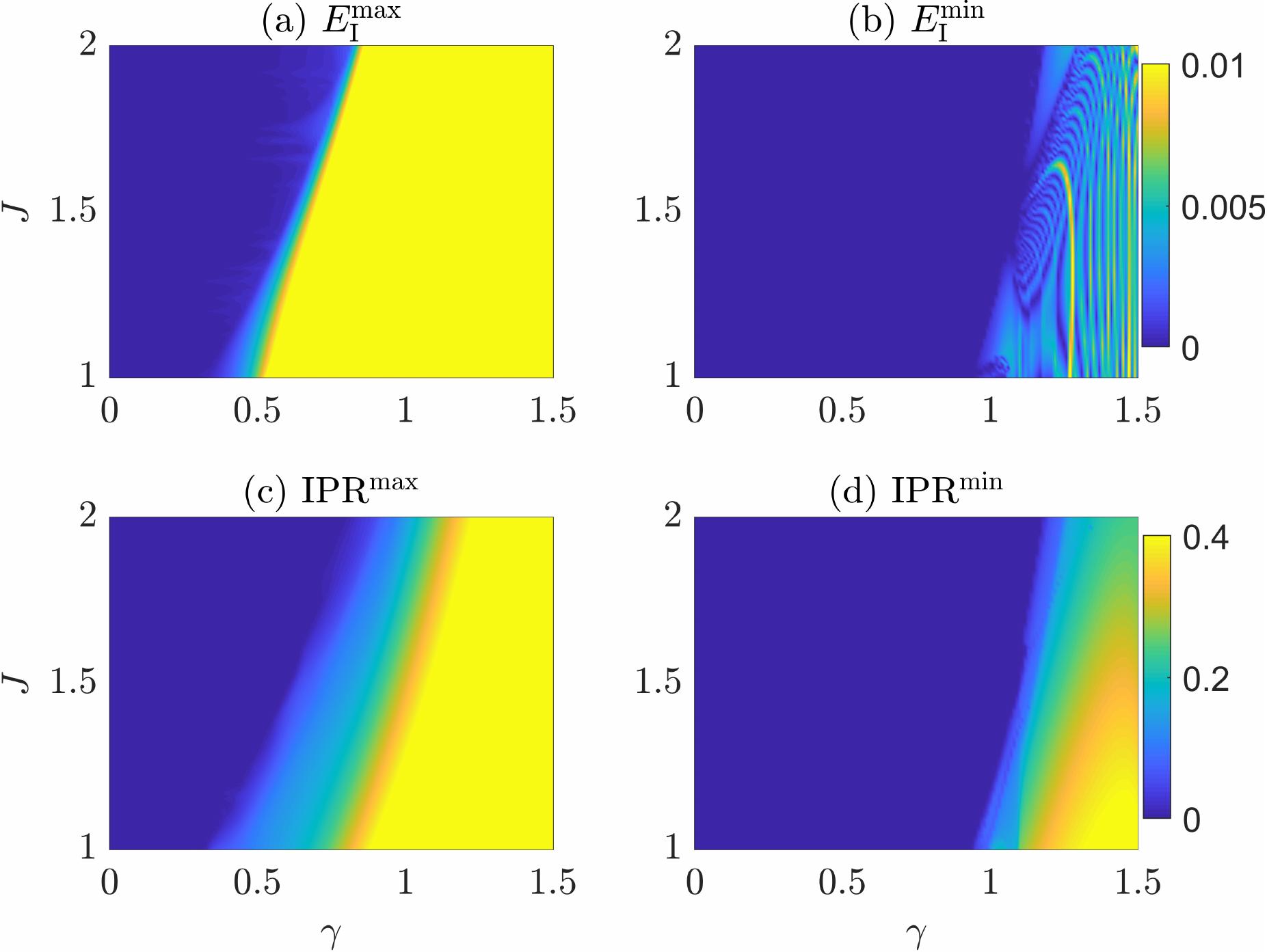}
		\par\end{centering}
	\caption{The maximum and minimum of the imaginary parts of eigenenergies and
		the IPRs versus the imaginary phase $\gamma$ and the hopping amplitude
		$J$ for the Model 3. Other system parameters are $V=0.5$, $\phi=\pi/2$
		and $\alpha=(\sqrt{5}-1)/2$ for all panels. The length of lattice
		is $L=610$. The panels (a) and (b) {[}(c) and (d){]} share the same
		color bar. \label{fig:M3-MaxMin-E-IPR}}
\end{figure}

In Fig.~\ref{fig:M3-MaxMin-E-IPR}, we report the extreme values of
${\rm Im}E$ {[}Eqs.~(\ref{eq:EImax}) and (\ref{eq:EImin}){]} and
IPRs {[}Eqs.~(\ref{eq:IPRmax}) and (\ref{eq:IPRmin}){]} versus the
imaginary non-Abelian phase $\gamma$ and the hopping amplitude $J$
for our Model 3. In Figs.~\ref{fig:M3-MaxMin-E-IPR}(a) and \ref{fig:M3-MaxMin-E-IPR}(c),
we observe that when the spectrum changes from real to complex across
the ${\cal PT}$-transition boundary, the ${\rm IPR}^{\max}$ also
increases from a vanishingly small value to a finite value at every given
$J$. Therefore, the ${\cal PT}$-breaking transition of the spectrum
happens together with the localization transition of the states between
extended and critical phases. We have also verified this observation
for different choices of the Abelian phase $\phi\in(-\pi,0)\cup(0,\pi)$.
Similarly, when the $E_{{\rm I}}^{\min}$ starts to deviate from zero
in Fig.~\ref{fig:M3-MaxMin-E-IPR}(b), all the eigenstates become
localized through a second transition as shown in Fig.~\ref{fig:M3-MaxMin-E-IPR}(d).
The energies of eigenstates are mostly complex after the this transition
from the critical to localized phases. Therefore, our non-Abelian
NHQC Model 3 could also possess extended, critical and localized phases
in a broad range of parameter domains. Moreover, the association between
non-Hermitian and non-Abelian potentials enables interesting patterns
of ${\cal PT}$-breaking and localization transitions among these
phases. Together with the previous two models, our results here uncovered
the rich physics that can be brought about by non-Abelian effects
in NHQCs.

\begin{figure}
	\begin{centering}
		\includegraphics[scale=0.47]{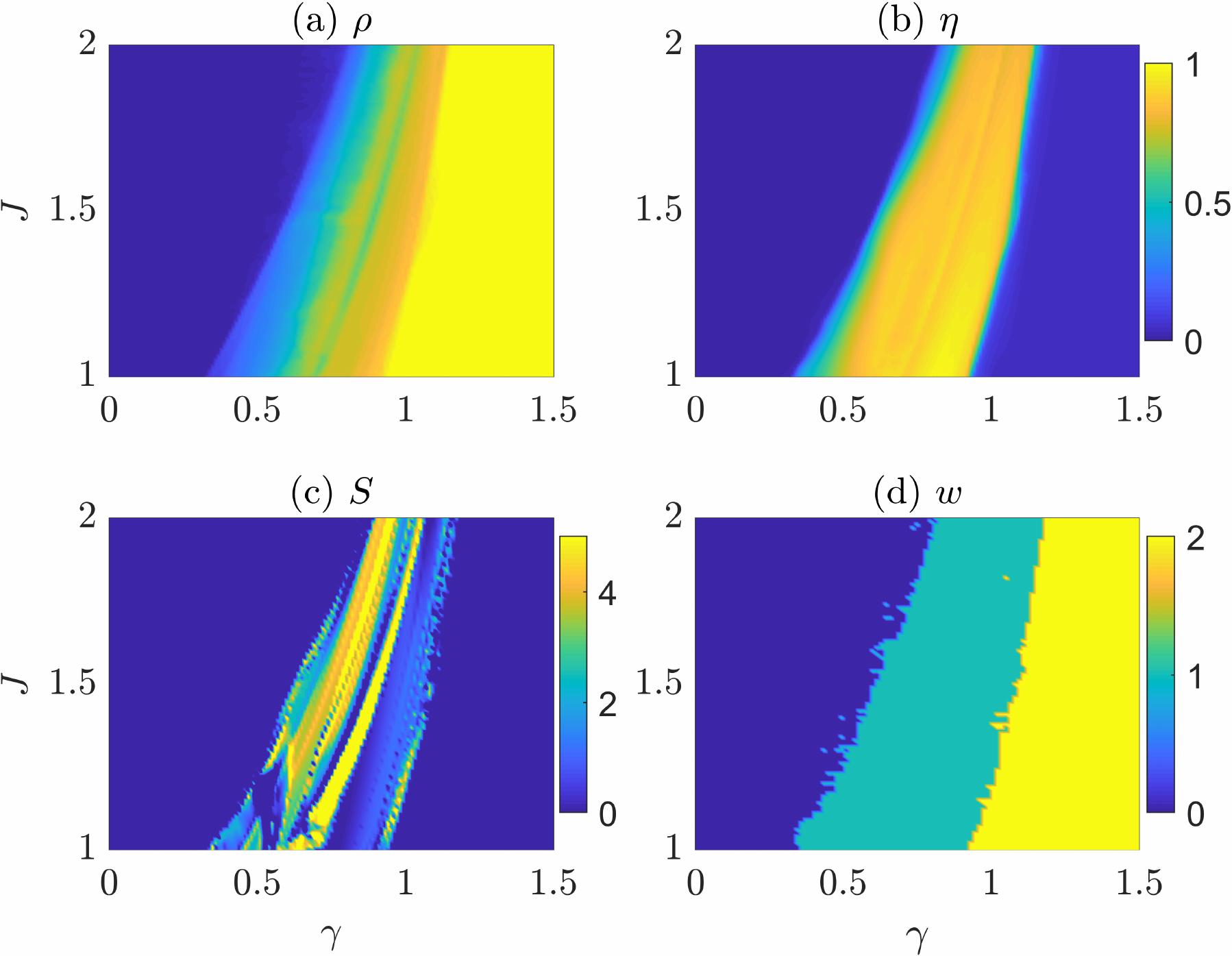}
		\par\end{centering}
	\caption{The density of states with real energies, the smoking-gun function
		of critical phase, the EE and the winding numbers versus the imaginary
		phase $\gamma$ and hopping amplitude $J$ for the Model 3 \cite{Note1}.
		Other system parameters are $V=0.5$, $\phi=\pi/2$ and $\alpha=(\sqrt{5}-1)/2$
		for all panels. The length of lattice is $L=610$. The panels (a)
		and (b) share the same color bar. \label{fig:M3-DOS-ETA-EE-W}}
\end{figure}

Finally, we present the the density of states with complex energies
{[}Eq.~(\ref{eq:DOS}){]}, the smoking-gun function of critical phases
{[}Eq.~(\ref{eq:ETA}){]}, the EE {[}Eq.~(\ref{eq:EE}){]} and the
phase diagram obtained from the winding numbers {[}Eq.~(\ref{eq:w12}){]}
of the Model 3 versus $\gamma$ and $J$ in Fig.~\ref{fig:M3-DOS-ETA-EE-W}.
We observe that in the regions with $\rho=0$ and $\rho\rightarrow1$
in Fig.~\ref{fig:M3-DOS-ETA-EE-W}(a), we also have ${\rm IPR}^{\max}\rightarrow0$
and ${\rm IPR}^{\min}>0$ in Figs.~\ref{fig:M3-MaxMin-E-IPR}(c) and
\ref{fig:M3-MaxMin-E-IPR}(d). These two regions thus correspond to
a ${\cal PT}$-invariant metallic phase with real spectrum and a ${\cal PT}$-broken
insulating phase with almost all eigenenergies being complex. The
region in between represents a critical phase with sufficient amounts
of coexisting real and complex eigenenergies in the spectrum ($0<\rho<1$).
This intermediate phase and its two boundaries are clearly highlighted
by the function $\eta$ in Fig.~\ref{fig:M3-DOS-ETA-EE-W}(b). In
Fig.~\ref{fig:M3-DOS-ETA-EE-W}(c), the EE is found to vanish ($S=0$)
both in the extended and localized phases. In the region of critical
mobility edge phase, the EE fluctuates strongly, as also observed
from the EE of the first two models. The phase diagram of Model 3
is shown in Fig.~\ref{fig:M3-DOS-ETA-EE-W}(d), where the blue, green
and yellow regions correspond to the extended, critical and localized
phases, respectively. The left and right boundaries are determined
by the locations where $w_{1}$ and $w_{2}$ get quantized jumps,
respectively. These boundaries are clearly coincident with the boundaries
of the ${\cal PT}$-symmetry breaking and the two localization transitions
identified from the results of $\rho$, ${\rm IPR}^{\max,\min}$ and
$S$. Therefore, both nonreciprocal hoppings and onsite gain/loss
could collaborate with non-Abelian quasiperiodic potentials to yield
different types of spectral transitions, topological localization
transitions and critical mobility edge phases in NHQCs.
Note in passing that the localization nature
of eigenstates in our models may also be characterized by
the energy-level statistics~\cite{NHLevelStat01,NHLevelStat02,NHLevelStat03,NHLevelStat04,NHLevelStat05,NHLevelStat06}.
We give brief accounts of this issue in the Appendix \ref{app1}.

\section{Conclusion and discussion\label{sec:Sum}}

In this work, we found unique phases and transitions that could be induced
by the interplay between non-Abelian potentials and different types
of non-Hermitian effects in 1D quasicrystals. For an NHQC
with either nonreciprocal hoppings or onsite gain and loss, we revealed
the emergence of a critical mobility edge phase when a non-Abelian
quasiperiodic modulation was introduced. Such a critical phase could be separated
from an extended phase and a localized phase by two topological localization
transitions. One of the transitions could further accompany a real-to-complex
spectral transition if the non-Abelian NHQC also possesses the ${\cal PT}$-symmetry.
We expect these features to be generic for any 1D non-Abelian NHQC models.
Our discoveries were further demonstrated by investigating the energy spectrum,
inverse participation ratios, entanglement spectrum, entanglement
entropy and spectral topological winding numbers for three prototypical
non-Abelian generalizations of non-Hermitian AAH models,
with each of them holding only a single ${\cal PT}$-breaking and
localization transition in the absence of non-Abelian effects. 
In short, we will have no ${\cal PT}$-breaking transitions and topological spectral windings in the Hermitian limits, no critical phases in the Abelian limits, and no localization transitions in the crystal limits ($\alpha\in{\mathbb Q}$) of our models. The rich physics we found are thus originated from the interplay among three key factors, i.e., the non-Hermitian effect, the quasiperiodic modulation, and the non-Abelian potential.
Our results thus enriched the study of NHQCs and
uncovered the intriguing phases and transitions that could be brought
about by non-Abelian effects.

The non-Abelian potentials considered in this work could be employed to induce critical phases in other 1D NHQCs. Qualitatively, the reason for the expansion of critical points in the Abelian limits of our models to critical phases in non-Abelian cases might be understood as follows. Let us consider the Eqs.~(\ref{eq:Seq})--(\ref{eq:d}) in the paper. Without the $d_n^x\sigma_x$ and $d_n^y\sigma_y$ terms in Eq.~(\ref{eq:PT0}), the Eq.~(\ref{eq:Seq}) can be decomposed into two spin polarized chains, which are described separately by
\begin{equation}
	J_{\rm L}\psi_{n+1,\uparrow}+J_{\rm R}\psi_{n-1,\uparrow}+V(d_n^0+d_n^z)\psi_{n,\uparrow}=E\psi_{n,\uparrow},\label{eq:chainup}
\end{equation}
\begin{equation}
	J_{\rm L}\psi_{n+1,\downarrow}+J_{\rm R}\psi_{n-1,\downarrow}+V(d_n^0-d_n^z)\psi_{n,\downarrow}=E\psi_{n,\downarrow}.\label{eq:chaindn}
\end{equation}
Referring to the Eq.~(\ref{eq:d}) of our three models, the Eqs.~(\ref{eq:chainup}) and (\ref{eq:chaindn}) describe two Abelian quasicrystals with the same critical point between extended and localized phases. There are no critical phases with mobility edges in these two decoupled chains. When the interchain coupling terms $d_n^x\sigma_x$ and $d_n^y\sigma_y$ in Eq.~(\ref{eq:PT0}) are switched on, the system in Eq.~(\ref{eq:Seq}) becomes genuine non-Abelian. Moreover, the terms $d_n^x\sigma_x+d_n^y\sigma_y$ may induce an overlap between the extended band of the spin-$\uparrow$ chain and the localized band of the spin-$\downarrow$ chain (or vice versa) through spin-flip couplings, leading to a critical phase with mobility edges. In other words, the non-Abelian interchain couplings split the two initially merged critical points of two decoupled chains and move them along opposite directions in the parameter space. The final result is the expansion of critical points of two originally disconnected chains to a critical phase of the non-Abelian coupled chain. Such a mechanism of generating critical phases may not be restricted to systems with non-Abelian potentials. It may also be used to understand the origin of critical phases in 1D systems with sublattice degrees of freedom or long-range hoppings. Nevertheless, regarding the interest of engineering non-Abelian potentials in different physical setups, our discoveries may provide further motivations for the realization of non-Abelian effects in systems beyond clean and Hermitian limits.

To the best of our knowledge, the ${\cal PT}$-breaking, localization and topological transitions have not been simultaneously explored in non-Abelian NHQCs. Our work thus introduced typical models and provided initial impetus along this line of study. It also has great experimental relevance regarding the recent realizations of NHQCs \cite{NHQC34,NHQC35}.
In experiments, NHQCs
were realized by nonunitary quantum walks in photonic
systems \cite{NHQC34,NHQC35}. Since an artificial spin-$1/2$ degree of freedom
is intrinsically present in the discrete time quantum walk, the existing
setups offer natural platforms to explore the critical phases, ${\cal PT}$-breaking
transitions and localization transitions in non-Abelian NHQCs.
The ${\cal PT}$-breaking transition may be revealed by measuring
the overall energy growth \cite{NHQC34} or the overall corrected probability \cite{NHQC35}
for an initially localized wavepacket in nonunitary quantum walks. Both quantities
will stay around their initial values in the $\cal{PT}$-invariant phase and grow over time
in the $\cal{PT}$-broken phase. The localization-delocalization transition may
be identified from the second moment of an initially localized wavepacket \cite{NHQC34},
which will show a monotonic increase (a low and bounded value) over time 
in the extended (localized) phase. The critical phase may be further located by measuring
both the dynamical IPR and NPR for a localized initial state \cite{NHQC35}. Both quantities
will deviate clearly from zero in the critical phase. Finally, a direct measurement of the
winding number could not be achieved in existing experimental setups. An indirect evidence 
for the winding number may be provided by the wave localization at a topological interface
between two distinct phases \cite{NHQC34}. For example, one may spatially connect a half-chain in the
extended phase with another half-chain in the localized phase at an interface site $j_0$.
A localized initial state in the extended half-chain will stop spreading and be trapped around $j_0$
when it reached the interface during the propagation. For the Models $1$ and $2$, the chiral transport
of wavepackets and non-Hermitian skin effects (NHSEs) may provide further evidences for nontrivial winding numbers \cite{NHQC35}.
In the Model 3, the hopping amplitudes are symmetric and no NHSEs are found, even though
there are spectral loops on the complex plane in the critical and localized phases (see Fig.~\ref{fig:M3-E-ES-IPR}).
In this case, we could not associate nontrivial winding numbers of the spectrum with NHSEs
and chiral propagations of wavepackets. More efforts are required to identify the
direct and generic connection bewteen the winding numbers and physical observables in non-Abelian NHQCs.

In future work, it is interesting to explore
NHQCs with other types of non-Abelian effects,
in superconducting systems, in quasi-one dimension \cite{Quasi1D} or higher
spatial dimensions and under time-periodic drivings \cite{NHFTI}.
Non-Hermitian localizations and mobility 
edges were also found in systems with random disorder \cite{NHDsod1,NHDsod2}.
Meanwhile, in similar models with correlated disorder, the critical phase might
be reduced to a critical point. The connection between the realness of an
eigenenergy and the localization nature of the related eigenstate could
also be different under random and correlated disorders~\cite{NHQC3,NHQC4}.
It would thus be interesting to further consider the impact of non-Abelian 
potentials on the spectrum, localization, topological 
and entanglement transitions in randomly disordered systems.
Beyond the single-particle case, possible many-body localizations and
critical phases originated from the competition between non-Abelian
quasiperiodic potentials and non-Hermitian effects also deserve to
be explored in detail.

\begin{acknowledgments}
	This work is supported by the National Natural Science Foundation of China (Grants No.~12275260 and 11905211), the Fundamental Research Funds for the Central Universities (Grant No.~202364008), and the Young Talents Project of Ocean University of China.
\end{acknowledgments}

\appendix

\section{Level statistics}\label{app1}
\begin{figure*}[ht]
	\begin{centering}
		\includegraphics[scale=0.88]{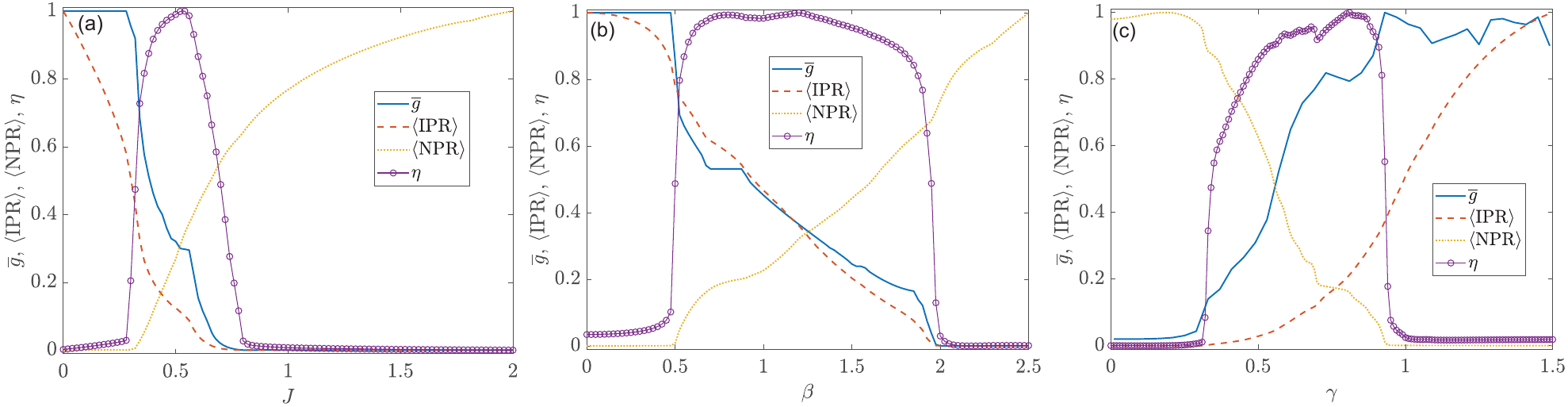}
		\par\end{centering}
	\caption{The averages of AGRs versus system parameters for the Models 1, 2, and 3 in (a), (b), and (c), respectively. Other system parameters are $(V,\phi)=(1,\pi/10)$ for (a), $(J,V,\phi)=(1,6,\pi/2)$ for (b), and $(J,V,\phi)=(1,0.5,\pi/2)$ for (c). We choose $\alpha=(\sqrt{5}-1)/2$ and $L=4181$ for all panels. The averages of IPRs, NPRs and the function $\eta$ [Eq.~(\ref{eq:ETA})] are also plotted to give a better illustration \cite{Note2}.}\label{fig:AGR}
\end{figure*}
In this Appendix, we discuss one type of tools based on the statistics of energy levels,
which may provide further signals about the phases with different localization nature
in our non-Abelian NHQCs. Let us denote the normalized
right eigenvectors of $H$ [Eq.~(\ref{eq:H})] and its eigenenergies
as $\{|\psi_{j}\rangle|j=1,...,N\}$ and $\{E_{j}|j=1,...,N\}$, with
$N$ being the total number of energy levels. Along
the real axis, the spacing between the $j$th and the $(j-1)$th levels
is given by $\epsilon_{j}={\rm Re}E_{j}-{\rm Re}E_{j-1}$, from which
we find the ratio between two adjacent spacings of energy levels as
$g_{j}=\min(\epsilon_{j},\epsilon_{j+1})/\max(\epsilon_{j},\epsilon_{j+1})$
for $j=2,...,N-1$.
Here the $\max(\epsilon_{j},\epsilon_{j+1})$ and $\min(\epsilon_{j},\epsilon_{j+1})$
yield the maximum and minimum between $\epsilon_{j}$ and $\epsilon_{j+1}$. 
The statistical property of adjacent gap ratios (AGRs) can
be obtained by averaging over all $g_{j}$ in the thermodynamic limit, i.e.,
\begin{equation}
	\overline{g}=\lim_{N\rightarrow\infty}\frac{1}{N}\sum_{j}g_{j}.\label{eq:AGR}
\end{equation}
We will have $\overline{g}\rightarrow0$ if all the bulk eigenstates
are extended. Comparatively, we expect $\overline{g}$
to approach a constant $\overline{g}_{\max}>0$ if all the bulk eigenstates
are localized. If $\overline{g}\in(0,\overline{g}_{\max})$,
extended and localized eigenstates should coexist and
the system should be in a critical phase. 
The $\overline{g}$ may thus be utilized to distinguish phases with different
localization nature in 1D non-Hermitian systems \cite{NHQC29}.

In Fig.~\ref{fig:AGR}, we report the averages of AGRs ${\overline g}$ for our three non-Abelian NHQC models in some typical situations. For the Models 1 and 2, we observe that the ${\overline g}$ indeed approaches zero and a finite constant $\overline{g}_{\max}$ in the extended and localized phases, respectively, and varying between them in the critical phase. For the Model 3, we also observe different tendencies for ${\overline g}$ in different phases. However, the ${\overline g}$ of Model 3 shows some oscillations in the localized phase, which indicate that the level statistics may have more complicated structures when the non-Abelian potential is also non-Hermitian.



\begin{thebibliography}{99}
	
	\bibitem{QC1} S. Aubry and G. Andr\'e, Analyticity
	breaking and Anderson localization in incommensurate lattices, Ann.
	Israel Phys. Soc \textbf{3}, 18 (1980).
	
	\bibitem{QC2} P. G. Harper, Single band motion of conduction electrons
	in a uniform magnetic field, Proc. Phys. Soc. London A \textbf{68},
	874 (1955).
	
	\bibitem{QC3} J. B. Sokoloff, Unusual band structure, wave function
	and electrical conductance in crystals with incommensurate periodic
	potentials, Phys. Rep. \textbf{126}, 189 (1985).
	
	\bibitem{QC4} B. Kramer and A. MacKinnon, Localization: Theory and
	experiment, Rep. Prog. Phys. \textbf{56}, 1469 (1993).
	
	\bibitem{QC5} A. Jagannathan, The Fibonacci quasicrystal: Case study
	of hidden dimensions and multifractality, Rev. Mod. Phys. \textbf{93},
	045001 (2021).
	
	\bibitem{NHQC0} P. Sarnak, Spectral behavior of quasi periodic potentials,
	Commun. Math. Phys. \textbf{84}, 377 (1982).
	
	\bibitem{NHQC1} A. Jazaeri and I. I. Satija, Localization transition
	in incommensurate non-Hermitian systems, Phys. Rev. E \textbf{63},
	036222 (2001).
	
	\bibitem{NHQC2} Q. Zeng, S. Chen, and R. L\"u, Anderson
	localization in the non-Hermitian Aubry-Andr\'e-Harper
	model with physical gain and loss, Phys. Rev. A \textbf{95}, 062118
	(2017).
	
	\bibitem{NHQC3} H. Jiang, L. Lang, C. Yang, S. Zhu, and S. Chen,
	Interplay of non-Hermitian skin effects and Anderson localization
	in nonreciprocal quasiperiodic lattices, Phys. Rev. B \textbf{100},
	054301 (2019).
	
	\bibitem{NHQC4} S. Longhi, Topological Phase Transition in non-Hermitian
	Quasicrystals, Phys. Rev. Lett. \textbf{122}, 237601 (2019).
	
	\bibitem{NHQC5} S. Longhi, Metal-insulator phase transition in a
	non-Hermitian Aubry-Andr\'e-Harper model, Phys. Rev.
	B \textbf{100}, 125157 (2019).
	
	\bibitem{NHQC6} T. Liu, H. Guo, Y. Pu, and S. Longhi, Generalized
	Aubry-Andr\'e self-duality and mobility edges in non-Hermitian
	quasiperiodic lattices, Phys. Rev. B \textbf{102}, 024205 (2020).
	
	\bibitem{NHQC7} Y. Liu, X.-P. Jiang, J. Cao, and S. Chen, Non-Hermitian
	mobility edges in one-dimensional quasicrystals with parity-time symmetry,
	Phys. Rev. B \textbf{101}, 174205 (2020).
	
	\bibitem{NHQC8} Q. Zeng, Y. Yang, and R. L\"u, Topological
	phases in one-dimensional nonreciprocal superlattices, Phys. Rev.
	B \textbf{101}, 125418 (2020).
	
	\bibitem{NHQC9} Q. Zeng, Y. Yang, and Y. Xu, Topological phases in
	non-Hermitian Aubry-Andr\'e-Harper models, Phys. Rev.
	B \textbf{101}, 020201(R) (2020).
	
	\bibitem{NHQC10} L. Zhai, S. Yin, and G. Huang, Many-body localization
	in a non-Hermitian quasiperiodic system, Phys. Rev. B \textbf{102},
	064206 (2020).
	
	\bibitem{NHQC11} Q. Zeng and Y. Xu, Winding numbers and generalized
	mobility edges in non-Hermitian systems, Phys. Rev. Research \textbf{2},
	033052 (2020).
	
	\bibitem{NHQC12} S. Longhi, Phase transitions in a non-Hermitian
	Aubry-Andr\'e-Harper model, Phys. Rev. B \textbf{103},
	054203 (2021).
	
	\bibitem{NHQC13} Y. Liu, Y. Wang, Z. Zheng, and S. Chen, Exact non-Hermitian
	mobility edges in one-dimensional quasicrystal lattice with exponentially
	decaying hopping and its dual lattice, Phys. Rev. B \textbf{103},
	134208 (2021).
	
	\bibitem{NHQC14} Y. Liu, Y. Wang, X. Liu, Q. Zhou, and S. Chen, Exact
	mobility edges, PT-symmetry breaking, and skin effect in one-dimensional
	non-Hermitian quasicrystals, Phys. Rev. B \textbf{103}, 014203 (2021).
	
	\bibitem{NHQC15} Z. Xu and S. Chen, Dynamical evolution in a one-dimensional
	incommensurate lattice with PT symmetry, Phys. Rev. A \textbf{103},
	043325 (2021).
	
	\bibitem{NHQC16} X. Cai, Boundary-dependent self-dualities, winding
	numbers, and asymmetrical localization in non-Hermitian aperiodic
	one-dimensional models, Phys. Rev. B \textbf{103}, 014201 (2021).
	
	\bibitem{NHQC17} L. Tang, G. Zhang, L. Zhang, and D. Zhang, Localization
	and topological transitions in non-Hermitian quasiperiodic lattices,
	Phys. Rev. A \textbf{103}, 033325 (2021).
	
	\bibitem{NHQC18} T. Liu, S. Cheng, H. Guo, and X. Gao, Fate of Majorana
	zero modes, exact location of critical states, and unconventional
	real-complex transition in non-Hermitian quasiperiodic lattices, Phys.
	Rev. B \textbf{103}, 104203 (2021).
	
	\bibitem{NHQC19} L. Zhai, G. Huang, and S. Yin, Cascade of the delocalization
	transition in a non-Hermitian interpolating Aubry-Andr\'e-Fibonacci
	chain, Phys. Rev. B \textbf{104}, 014202 (2021).
	
	\bibitem{NHQC20} L. Zhou, Floquet engineering of topological localization
	transitions and mobility edges in one-dimensional non-Hermitian quasicrystals,
	Phys. Rev. Research \textbf{3}, 033184 (2021).
	
	\bibitem{NHQC21} L. Zhou and W. Han, Non-Hermitian quasicrystal in
	dimerized lattices, Chinese Phys. B \textbf{30}, 100308 (2021).
	
	\bibitem{NHQC22} Z.-H. Wang, F. Xu, L. Li, D. Xu, and B. Wang, Unconventional
	real-complex spectral transition and Majorana zero modes in nonreciprocal
	quasicrystals, Phys. Rev. B \textbf{104}, 174501 (2021).
	
	\bibitem{NHQC23} Y. Liu, Q. Zhou, and S. Chen, Localization transition,
	spectrum structure, and winding numbers for one-dimensional non-Hermitian
	quasicrystals, Phys. Rev. B \textbf{104}, 024201 (2021).
	
	\bibitem{NHQC24} X. Cai, Localization and topological phase transitions
	in non-Hermitian Aubry-Andr\'e-Harper models with p-wave
	pairing, Phys. Rev. B \textbf{103}, 214202 (2021).
	
	\bibitem{NHQC25} S. Longhi, Non-Hermitian Maryland model, Phys. Rev.
	B \textbf{103}, 224206 (2021).
	
	\bibitem{NHQC26} A. P. Acharya, A. Chakrabarty, and D. K. Sahu, Localization,
	PT-Symmetry Breaking and Topological Transitions in non-Hermitian
	Quasicrystals, Phys. Rev. B \textbf{105}, 014202 (2022).
	
	\bibitem{NHQC27} C. Yuce and H. Ramezani, Coexistence of extended
	and localized states in the one-dimensional non-Hermitian Anderson
	model, Phys. Rev. B \textbf{106}, 024202 (2022).
	
	\bibitem{NHQC28} L. Zhou and Y. Gu, Topological delocalization transitions
	and mobility edges in the nonreciprocal Maryland model, J. Phys.:
	Condens. Matter \textbf{34}, 115402 (2022).
	
	\bibitem{NHQC29} W. Han and L. Zhou, Dimerization-induced mobility
	edges and multiple reentrant localization transitions in non-Hermitian
	quasicrystals, Phys. Rev. B \textbf{105}, 054204 (2022).
	
	\bibitem{NHQC30} L. Zhou and W. Han, Driving-induced multiple PT-symmetry
	breaking transitions and reentrant localization transitions in non-Hermitian
	Floquet quasicrystals, Phys. Rev. B \textbf{106}, 054307 (2022).
	
	\bibitem{NHQC31} X. Xia, K. Huang, S. Wang, and X. Li, Exact mobility
	edges in the non-Hermitian $t_{1}$-$t_{2}$ model: Theory and possible
	experimental realizations, Phys. Rev. B \textbf{105}, 014207 (2022).
	
	\bibitem{NHQC32} T. Liu and X. Xia, Real-complex transition driven
	by quasiperiodicity: A class of non-PT symmetric models, Phys. Rev.
	B \textbf{105}, 054201 (2022).
	
	\bibitem{NHQC33} L. Zhai, G. Huang, and S. Yin, Nonequilibrium dynamics
	of the localization-delocalization transition in the non-Hermitian
	Aubry-Andr\'e model, Phys. Rev. B \textbf{106}, 014204
	(2022).
	
	\bibitem{NHQC34} S. Weidemann, M. Kremer, S. Longhi and A. Szameit,
	Topological triple phase transition in non-Hermitian Floquet quasicrystals,
	Nature \textbf{601}, 354-359 (2022).
	
	\bibitem{NHQC35} Q. Lin, T. Li, L. Xiao, K. Wang, W. Yi, and P. Xue,
	Topological Phase Transitions and Mobility Edges in Non-Hermitian
	Quasicrystals, Phys. Rev. Lett. \textbf{129}, 113601 (2022).
	
	\bibitem{NHQC36} S. Longhi, Non-Hermitian topological mobility edges
	and transport in photonic quantum walks, Opt. Lett. \textbf{47}, 2951
	(2022).
	
	\bibitem{NHQC37} X. Cai, Localization transitions and winding numbers
	for non-Hermitian Aubry-Andr\'e-Harper models with
	off-diagonal modulations, Phys. Rev. B \textbf{106}, 214207 (2022).
	
	\bibitem{NHQC38} W. Chen, S. Cheng, J. Lin, R. Asgari, and X. Gao,
	Breakdown of the correspondence between the real-complex and delocalization-localization
	transitions in non-Hermitian quasicrystals, Phys. Rev. B \textbf{106},
	144208 (2022).
	
	\bibitem{NHQC39} L.-M. Chen, Y. Zhou, S. A. Chen, and P. Ye, Quantum
	entanglement of non-Hermitian quasicrystals, Phys. Rev. B \textbf{105},
	L121115 (2022).
	
	\bibitem{NAGPRev1} J. Dalibard, F. Gerbier, G. Juzeli\=unas,
	and P. \"Ohberg, Colloquium: Artificial gauge potentials
	for neutral atoms, Rev. Mod. Phys. \textbf{83}, 1523 (2011).
	
	\bibitem{NAGPRev2} M. Lewenstein, A. Sanpera, and V. Ahufinger, \emph{Ultracold
		Atoms in Optical Lattices} (Oxford University Press, Oxford, United
	Kingdom, 2012).
	
	\bibitem{NAGPRev3} U.-J. Wiese, Ultracold quantum gases and lattice
	systems: quantum simulation of lattice gauge theories, Ann. Phys.
	(Berlin) \textbf{525}, 777-796 (2013).
	
	\bibitem{NAGPRev4} N. Goldman, G. Juzeli\=unas,
	P. \"Ohberg, and I. B. Spielman, Light-induced gauge
	fields for ultracold atoms, Rep. Prog. Phys. \textbf{77}, 126401 (2014).
	
	\bibitem{NAGPRev5} N. Goldman, J. C. Budich, and P. Zoller, Topological
	quantum matter with ultracold gases in optical lattices, Nat. Phys.
	\textbf{12}, 639-645 (2016).
	
	\bibitem{NAGPRev6} M. Aidelsburger, S. Nascimbene, and N. Goldman,
	Artificial gauge fields in materials and engineered systems, C. R.
	Physique \textbf{19}, 394-432 (2018).
	
	\bibitem{NAGPRev7} Y. Han, W. Yi, and W. Zhang, \emph{Physics on
		Ultracold Quantum Gases} (World Scientific Press, Singapore, 2019).
	
	\bibitem{NAGPRev8} H. Zhai, \emph{Ultracold Atomic Physics} (Cambridge
	University Press, Cambridge, United Kingdom, 2021).
	
	\bibitem{Hofstadter} D. Hofstadter, Energy levels and wave functions
	of Bloch electrons in rational and irrational magnetic fields, Phys.
	Rev. B \textbf{14}, 2239 (1976).
	
	\bibitem{NAHH1} K. Osterloh, M. Baig, L. Santos, P. Zoller, and M.
	Lewenstein, Cold Atoms in Non-Abelian Gauge Potentials: From the Hofstadter
	``Moth'' to Lattice Gauge Theory, Phys. Rev. Lett. \textbf{95},
	010403 (2005).
	
	\bibitem{NAHH2} I. I. Satija, D. C. Dakin, and C. W. Clark, Metal-Insulator
	Transition Revisited for Cold Atoms in Non-Abelian Gauge Potentials,
	Phys. Rev. Lett. \textbf{97}, 216401 (2006).
	
	\bibitem{NAHH3} I. I. Satija, D. C. Dakin, J. Y. Vaishnav, and C.
	W. Clark, Physics of a two-dimensional electron gas with cold atoms
	in non-Abelian gauge potentials, Phys. Rev. A \textbf{77}, 043410
	(2008).
	
	\bibitem{NAHH4} J. Wang and J. Gong, Butterfly Floquet Spectrum in
	Driven SU(2) Systems, Phys. Rev. Lett. \textbf{102}, 244102 (2009).
	
	\bibitem{NAHH5} N. Goldman, A. Kubasiak, P. Gaspard, and M. Lewenstein,
	Ultracold atomic gases in non-Abelian gauge potentials: The case of
	constant Wilson loop, Phys. Rev. A \textbf{79}, 023624 (2009).
	
	\bibitem{NAHH6} N. Goldman, I. Satija, P. Nikolic, A. Bermudez, M.
	A. Martin-Delgado, M. Lewenstein, and I. B. Spielman, Phys. Rev. Lett.
	\textbf{105}, 255302 (2010).
	
	\bibitem{NAHH7} A. Kosior and K. Sacha, Simulation of non-Abelian
	lattice gauge fields with a single-component gas, EPL (Europhysics
	Letters) \textbf{107}, 26006 (2014).
	
	\bibitem{NAHH8} Y. Li, Time-reversal invariant SU(2) Hofstadter problem
	in three-dimensional lattices, Phys. Rev. B \textbf{91}, 195133 (2015).
	
	\bibitem{NAHH9} J.-Q. Cai, Q.-Y. Yang, Z.-Y. Xue, M. Gong, G.-C.
	Guo, and Y. Hu, Interplay between non-Hermiticity and non-Abelian
	gauge potential in topological photonics, arXiv:1812.02610.
	
	\bibitem{NAHH10} Y. Yang, C. Peng, D. Zhu, H. Buljan, J. D. Joannopoulos,
	B. Zhen, and M. Solja{\v c}i\'c, Synthesis
	and observation of non-Abelian gauge fields in real space, Science
	\textbf{365}, 1021-1025 (2019).
	
	\bibitem{NAHH11} E. G. Guan, H. Yu, and G. Wang, Non-Abelian gauge
	potentials driven localization transition in quasiperiodic optical
	lattices. Phys. Lett. A \textbf{384}, 126152 (2020).
	
	\bibitem{NAHH12} Y. Yang, B. Zhen, J. D. Joannopoulos, and M. Solja{\v c}i\'c,
	Non-Abelian generalizations of the Hofstadter model: spin-orbit-coupled
	butterfly pairs, Light Sci. Appl. \textbf{9}, 177 (2020).
	
	\bibitem{NAHH13} V. Liu, Y. Yang, J. D. Joannopoulos, and M. Solja{\v c}i\'c,
	Three-dimensional non-Abelian generalizations of the Hofstadter model:
	Spin-orbit-coupled butterfly trios, Phys. Rev. B \textbf{104}, 115127
	(2021).
	
	\bibitem{NAHH14} D. Cheng, K. Wang, and S. Fan, Artificial non-Abelian
	lattice gauge fields for photons in the synthetic frequency dimension,
	Phys. Rev. Lett. {\bf 130}, 083601 (2023).
	
	\bibitem{MCMP} S. M. Girvin and K. Yang, \emph{Modern Condensed Matter
		Physics} (Cambridge University Press, Cambridge, United Kingdom, 2019).
	
	\bibitem{Relocal} S. Roy, T. Mishra, B. Tanatar, and S. Basu, Reentrant
	Localization Transition in a Quasiperiodic Chain, Phys. Rev. Lett.
	\textbf{126}, 106803 (2021).
	
	\bibitem{FloqESEE} L. Zhou, Entanglement spectrum and entropy in
	Floquet topological matter, Phys. Rev. Research \textbf{4}, 043164
	(2022).
	
	\bibitem{Peschel2003} I. Peschel, Calculation of reduced density
	matrices from correlation functions, J. Phys. A: Math. Gen. \textbf{36},
	L205 (2003).
	
	\bibitem{NHESEE1} L. Herviou, J. H. Bardarson, and N. Regnault, Defining
	a bulk-edge correspondence for non-Hermitian Hamiltonians via singular-value
	decomposition, Phys. Rev. A \textbf{99}, 052118 (2019).
	
	\bibitem{NHESEE2} L. Herviou, N. Regnault, and J. H. Bardarson, Entanglement
	spectrum and symmetries in non-Hermitian fermionic non-interacting
	models, SciPost Phys. \textbf{7}, 069 (2019).
	
	\bibitem{NHESEE3} P.-Y. Chang, J.-S. You, X. Wen, and S. Ryu, Entanglement
	spectrum and entropy in topological non-Hermitian systems and nonunitary
	conformal field theory, Phys. Rev. Research \textbf{2}, 033069 (2020).
	
	\bibitem{NHESEE4} S. Mu, C. H. Lee, L. Li , and J. Gong, Emergent
	Fermi surface in a many-body non-Hermitian fermionic chain, Phys.
	Rev. B \textbf{102}, 081115(R) (2020).
	
	\bibitem{NHESEE5} L.-M. Chen, S. A. Chen, and P. Ye, Entanglement,
	non-hermiticity, and duality, SciPost Phys. \textbf{11}, 3 (2021).
	
	\bibitem{NHESEE6} Y.-B. Guo, Y.-C. Yu, R.-Z. Huang, L.-P. Yang, R.-Z.
	Chi, H.- J. Liao, and T. Xiang, Entanglement entropy of non-Hermitian
	free fermions, J. Phys.: Condens. Matter \textbf{33}, 475502 (2021).
	
	\bibitem{NHESEE7} N. Okuma and M. Sato, Quantum anomaly, non-Hermitian
	skin effects, and entanglement entropy in open systems, Phys. Rev.
	B \textbf{103}, 085428 (2021).
	
	\bibitem{NHESEE8} C. Ortega-Taberner. L. R{\o}dland,
	and M. Hermanns, Polarization and entanglement spectrum in non-Hermitian
	systems, Phys. Rev. B \textbf{105}, 075103 (2022).
	
	\bibitem{Note1} We rescaled some quantities in
	order for a more compact illustration. The presented
	$E_{{\rm I}}^{\max}$ is rescaled by its maximum over the considered
	domain of system parameters in the figure. The presented $\eta$ is
	shifted and rescaled as $[\eta-\min(\eta)]/\max[\eta-\min(\eta)]$,
	where the $\min(\eta)$ and $\max(\eta)$ refer to the minimum and
	maximum of $\eta$ over the considered domain of system parameters
	in the figure. Both the values of the presented $E_{{\rm I}}^{\max}$
	and $\eta$ are thus confined in the range $[0,1]$.
	
	\bibitem{Quasi1D} S. Mu, L. Zhou, L. Li, and J. Gong, Non-Hermitian
	pseudo mobility edge in a coupled chain system, Phys. Rev. B \textbf{105},
	205402 (2022).
	
	\bibitem{NHFTI} L. Zhou and J. Gong, Non-Hermitian Floquet topological
	phases with arbitrarily many real-quasienergy edge states, Phys. Rev.
	B \textbf{98}, 205417 (2018).
	
	\bibitem{NHLevelStat01} I. Y. Goldsheid and B. A. Khoruzhenko, Distribution of Eigenvalues in Non-Hermitian Anderson Models, Phys. Rev. Lett. {\bf 80}, 2897 (1998).
	
	\bibitem{NHLevelStat02} J. T. Chalker and B. Mehlig, Eigenvector Statistics in Non-Hermitian Random Matrix Ensembles, Phys. Rev. Lett. {\bf 81}, 3367 (1998).
	
	\bibitem{NHLevelStat03} L. G. Molinari, Non-Hermitian spectra and Anderson localization, J. Phys. A: Math. Theor. {\bf 42}, 265204 (2009).
	
	\bibitem{NHLevelStat04} R. Hamazaki, K. Kawabata, N. Kura, and M. Ueda, Universality classes of non-Hermitian random matrices, Phys. Rev. Research {\bf 2}, 023286 (2020).
	
	\bibitem{NHLevelStat05} L. S\'a, P. Ribeiro, and T. Prosen, Complex Spacing Ratios: A Signature of Dissipative Quantum Chaos, Phys. Rev. X {\bf 10}, 021019 (2020).
	
	\bibitem{NHLevelStat06} S. Ghosh, S. Gupta, and M. Kulkarni, Spectral properties of disordered interacting non-Hermitian systems, Phys. Rev. B {\bf 106}, 134202 (2022).
	
	\bibitem{NHDsod1} N. Hatano and D. R. Nelson, Localization Transitions in Non-Hermitian Quantum Mechanics, Phys. Rev. Lett. {\bf 77}, 570 (1996).
	
	\bibitem{NHDsod2} J. Feinberg and A. Zee, Non-Hermitian localization and delocalization, Phys. Rev. E {\bf 59}, 6433 (1999).
	
	\bibitem{Note2} We have scaled all the quantities to restrict their ranges to $[0,1]$ (see also \cite{Note1}). The ${\overline g}$, $\langle{\rm IPR}\rangle$ and $\langle{\rm NPR}\rangle$ are scaled by the maximum of AGRs, IPRs and NPRs at each $J$, $\beta$ and $\gamma$ for the three models. The non-scaled values of ${\overline g}$ are approximately $0.44$, $0.49$ and $0.41$ for the Models $1$, $2$ and $3$ in their localized phases.
	 
	
\end{thebibliography}
\end{document}